\documentclass{aa}

\usepackage[utf8]{inputenc}
\usepackage{graphicx}
\usepackage[colorlinks=true, allcolors=blue]{hyperref}
\usepackage{xcolor}
\usepackage{txfonts}
\makeatletter
\renewcommand*\aa@pageof{, page \thepage{} of \pageref*{LastPage}}
\makeatother

\usepackage{multirow}
\usepackage{amsmath}
\usepackage{comment}

\begin{document}

\date{}

\title{Hertzsprung gap stars in nearby galaxies and the Quest for Luminous Red Novae Progenitors\thanks{Tables 3 and 4 are only available in electronic form
at the CDS via anonymous ftp to cdsarc.cds.unistra.fr (130.79.128.5)
or via https://cdsarc.cds.unistra.fr/cgi-bin/qcat?J/A+A/}}
\author{Hugo Tranin\inst{\ref{iccub}, \ref{fqa},\ref{ieec}}, Nadejda Blagorodnova\inst{\ref{iccub}, \ref{fqa}, \ref{ieec}}, Viraj Karambelkar\inst{\ref{caltech}}, Paul J. Groot\inst{\ref{radboud}, \ref{capetown}, \ref{saao},\ref{idia}}, Steven Bloemen\inst{\ref{radboud}}, Paul M. Vreeswijk\inst{\ref{radboud}}, Dani\"elle L.A. Pieterse\inst{\ref{radboud}}, and Jan van Roestel\inst{\ref{uam}}}

\institute{Institut de Ciències del Cosmos (ICCUB), Universitat de Barcelona (UB), c. Martí i Franquès, 1, 08028, Barcelona, Spain
e-mail: \texttt{hugo.tranin@irap.omp.eu} \label{iccub}
\and
Departament de Física Quàntica i Astrofísica (FQA), Universitat de Barcelona (UB), c. Martí i Franquès, 1, 08028, Barcelona, Spain
\label{fqa}
\and
Institut d'Estudis Espacials de Catalunya (IEEC), c. Gran Capità, 2-4, 08034, Barcelona, Spain
\label{ieec}
\and
Cahill Center for Astrophysics, California Institute of Technology, Pasadena, CA 91125, USA
\label{caltech}
\and
Department of Astrophysics/IMAPP, Radboud University, PO Box 9010, 6500 GL Nijmegen, The Netherlands
\label{radboud}
\and
Department of Astronomy, University of Cape Town, Private Bag
X3, Rondebosch, 7701, South Africa
\label{capetown}
\and
South African Astronomical Observatory, P.O. Box 9, Observatory,
7935, South Africa
\label{saao}
\and
The Inter-University Institute for Data Intensive Astronomy, University of Cape Town, Private Bag X3, Rondebosch, 7701, South Africa
\label{idia}
\and
Anton Pannekoek Institute for Astronomy, University of Amsterdam, P.O. Box 94249, 1090 GE Amsterdam, The Netherlands
\label{uam}
}

\abstract
    {After the main sequence phase, stars more massive than 2.5 M$_\odot$ rapidly evolve through the Hertzsprung gap as yellow giants and supergiants (YSG), before settling into the red giant branch. Identifying Hertzsprung gap stars in nearby galaxies is crucial for pinpointing progenitors of luminous red novae (LRNe) -- astrophysical transients attributed to stellar mergers. In the era of extensive transient surveys like the Vera Rubin Observatory's Legacy Survey of Space and Time (LSST), this approach offers a new way to predict and select common envelope transients.}
    {This study investigates potential progenitors and precursors of LRNe by analysing \textit{Hubble Space Telescope} (HST) photometry of stellar populations in galaxies within $\sim$20 Mpc to identify YSG candidates. Additionally, we use the Zwicky Transient Facility (ZTF) and MeerLICHT/BlackGEM to identify possible precursors, preparing for future observations by the LSST.}
    {We compiled a sample of 369 galaxies with HST exposures in the $F475W$, $F555W$, $F606W$, and $F814W$ filters. We identified YSG candidates using MESA stellar evolution tracks and statistical analysis of colour-magnitude diagrams (CMDs).}
    {Our sample includes 154,494 YSG candidates with masses between $3 M_\odot$ and $20 M_\odot$ and is affected by various contaminants, notably foreground stars and extinguished main-sequence stars. After excluding foreground stars using \textit{Gaia} proper motions, contamination is estimated at 1\% from foreground stars (based on TRILEGAL simulations) and $\sim$20\% from extinction affecting main-sequence stars. Combining our YSG candidates with time-domain catalogues yielded several interesting candidates. In particular, we identified twelve LRN precursor candidates for which followup is encouraged.}
    {We highlight the importance of monitoring future transients that match YSG candidates to avoid missing potential LRNe and other rare transients. LSST will be a game changer in the search for LRN progenitors and precursors, discovering over 300,000 new YSG candidates and 100 LRN precursors within 20 Mpc.}
    {}
\titlerunning{Quest for Hertzsprung gap stars and LRNe precursors}
\authorrunning{H. Tranin et al.}

\maketitle

\section{Introduction}
\label{sec:intro}

The termination of hydrogen burning in a stellar core marks the end of a star's main sequence (MS) phase. As the core contracts and temperatures rise, hydrogen shell burning starts, causing the star to expand rapidly. This expansion leads to the star becoming progressively larger and cooler, transitioning into the yellow giant (YG) and supergiant (YSG) phase. However, this evolutionary stage is brief and difficult to observe, creating a "gap" between the MS and the Red Giant Branch (RGB) in the Hertzsprung-Russell (HR) diagram, known as the Hertzsprung gap. Early surveys of YSGs in the Milky Way, Magellanic Clouds, and Andromeda revealed significant discrepancies between observations and stellar evolution models, particularly in estimating the duration of this phase \citep{Massey2000, Drout2009, Neugent2010, Drout2012}.

Adding further complexity, most massive stars are found in binary or multiple systems, with 70\% of O-type stars and 50\% of B-type stars having companions (e.g., \citealt{Moe2017ApJS} and references therein). The rapid expansion of YSGs in such systems often triggers interactions like Case B mass transfer and substantial mass loss \citep{Marchant2023}. This interaction makes the YSG phase particularly valuable for studying binary evolution. 
In cases where a YSG is part of a binary system, unstable mass transfer can lead to a common envelope phase \citep{Paczynski1976IAUS}. The transfer of angular momentum of the binary into the envelope can result in its partial (or total) ejection, producing a rare astrophysical transient known as a Luminous Red Nova (LRN). While YSGs have been linked to other rare transients, such as core-collapse supernovae (SNe) \citep{Smartt2015} and failed SNe \citep{Georgy2012, Neustadt2021}, they are most notably identified as progenitors of LRNe \citep{MacLeod2017,Blagorodnova2017,Blagorodnova2021,Cai2022}.


LRNe are optical and infrared transients with luminosities between those of novae and supernovae, evolving over several weeks to months. Their peak brightness correlates with the progenitor's mass \citep{Kochanek2014,Blagorodnova2021}, enabling us to investigate binary evolution across a wide range of stellar masses, from low-mass to massive stars. Unlike optical LRNe, infrared LRNe may result from common envelope ejections in stars more evolved than YSGs \citep{MacLeod2022}. These stars experience significant mass loss, causing them to become obscured in optical wavelengths while remaining detectable in the infrared. The most well-studied LRN to date, V1309 Sco, was discovered in the Milky Way. Extensive photometric data up to seven years before the onset of the transient revealed its progenitor, a contact binary system with a quickly decaying period, possibly evolving to the merger of its components \citep{Tylenda2011}.
Interestingly, the light curve of V1309 Sco exhibited a slow, steady brightening beginning approximately five years prior to the merger. Similar precursor emissions have been observed in extragalactic LRNe \citep{Kankare2015, Blagorodnova2017, Blagorodnova2020, Pastorello2019b, Pastorello2021}, though their greater distances and faintness have made it difficult to achieve equally detailed sampling. 
It remains an open question whether this precursor phase reflects continuous or episodic mass loss in the years leading up to the major outburst, potentially signaling the early stages of a binary system merger. Long-term observations of LRN progenitor stars are therefore crucial for testing their binary origins and gaining valuable insights into pre-merger mass loss episodes in these systems.


The primary challenge in detecting precursors is their modest absolute magnitude, which means most extragalactic LRNe are only identified during their brightest outbursts. The lack of deep archival data further complicates efforts to detect precursor brightening. To overcome this, a comprehensive catalog of potential progenitors in nearby galaxies is essential. A large and deep survey is mandatory to achieve such a goal.
While early discoveries of LRNe were limited to the Milky Way and M31 (e.g., V4332 Sgr and V838 Mon, \citealt{Martini1999, Tylenda2006}; M31 LRN-2015, \citealt{Williams2015, Blagorodnova2020}), the advent of large synoptic surveys, such as ATLAS \citep{Tonry2018PASP}, the Zwicky Transient Facility \citep[ZTF;][]{Bellm2019}, MeerLICHT \citep{Bloemen2016}, and BlackGEM \citep{Groot2019,Groot2024}, has paved the way for the systematic detection of LRNe throughout the Local Group and beyond  (e.g. \citealt{Karambelkar2023}). The upcoming Vera C. Rubin Observatory, with its Legacy Survey of Space and Time (LSST), promises to revolutionize transient astronomy, providing an unprecedented opportunity to detect faint LRNe and LRN precursors across vast distances. With LSST expected to generate 10 million transient alerts daily, the primary challenge will be identifying LRN precursors amidst this vast data unless a targeted strategy is developed to detect them. 

In this study, we present a strategy for identifying LRN precursors by cataloging YSG candidates within 20 Mpc, leveraging archival \textit{Hubble Space Telescope} (HST) photometry. Our goal is to identify potential precursor candidates by cross-matching future transient alerts with this catalog, offering a new approach to studying the processes driving LRN precursor brightening and triggering mechanisms. 
An initial study provided a similar census of YGs and YSGs in the Milky Way \citep{Addison2022}. Using \textit{Gaia} DR2 and EDR3 \citep{Gaia2018, GaiaEDR3} sources, they modelled the distribution of a sample of Milky Way stars in the colour-magnitude ($M_G$ vs. $BP-RP$) diagram. Statistical modelling allowed them to select the Hertzsprung gap as the least-populated region of the diagram. 
As a result, they identified 21 candidates exhibiting signs of a steady increase in brightness. Most of these candidates showed Balmer lines in emission and an infrared excess in their spectral energy distributions, reinforcing the likelihood of an accreting system with dust production. Interestingly, one of them turned out to be the precursor of a type-I X-ray burst 
\citep{ATel2022}. 

While the census of Galactic YSG candidates is crucial, the rate of LRNe is comparable to that of core-collapse supernovae \citep{Karambelkar2023}, suggesting that the largest population of LRN precursors may be found beyond our Galaxy. To address this, we apply a similar methodology to catalog extragalactic YSGs using archival HST photometry of galaxies within 20 Mpc. If a future transient alert matches an object in this catalog, it could be flagged as a potential precursor and prioritized for follow-up observations. This approach offers a practical strategy to sift through the vast number of LSST alerts and conduct the first in-depth study of the processes driving LRN precursor brightening.

The paper is structured as follows: in Section \ref{sec:method} we define the data and methods used to select the samples and remove contaminants. Section \ref{sec:results} presents a statistical description of the selected sample and the different sources of contamination. We crossmatched this catalogue to LRN progenitors from the literature, to the Transient Name Server\footnote{\url{https://www.wis-tns.org/}} (TNS) and to time-domain surveys in Section \ref{sec:application}, showcasing some individual precursor candidates. These results are discussed in Section \ref{sec:discussion} and compared to previous searches for YSGs and LRN progenitors. We conclude and give some guidance on how to use this catalogue in Section \ref{sec:conclusion}.

\section{Methods}\label{sec:method}
\subsection{Selection criteria}


To identify progenitors of transients produced by massive stars, we need a YSG sample complete down to 8 $M_\odot$. This completeness is key for future population studies. According to MESA (Modules for Experiments in Stellar Astrophysics, \citealt{Paxton2011}) stellar models, an 8 $M_\odot$ star crossing the Hertzsprung gap has an absolute magnitude of $M_{F814W} \sim -5$. HST can detect such a star up to 20 Mpc away with a 1-hour exposure ($m_{lim,F814W} \sim 27.5 $). Shorter typical exposures (e.g., 600s) only detect YSGs down to 16 $M_\odot$ at this distance. Thus, our sample is limited to galaxies within 20 Mpc with deep HST exposures in several bands,  allowing us to conduct a census of stars in the Local Universe up to the Virgo cluster, and ensuring a minimum completeness limit of 16 $M_\odot$ to cover the Hertzsprung gap.

\begin{figure}
    \centering
    \includegraphics[width=8.5cm]{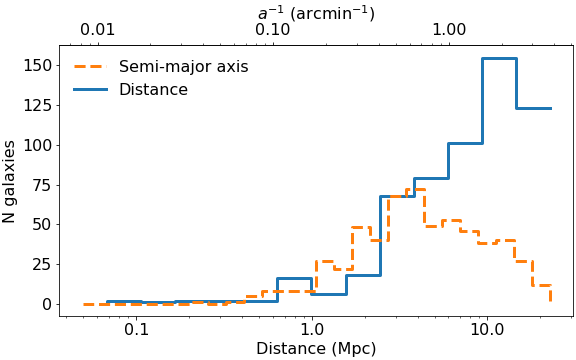}

\vspace{0.2 cm}
     \includegraphics[width=8.5cm]{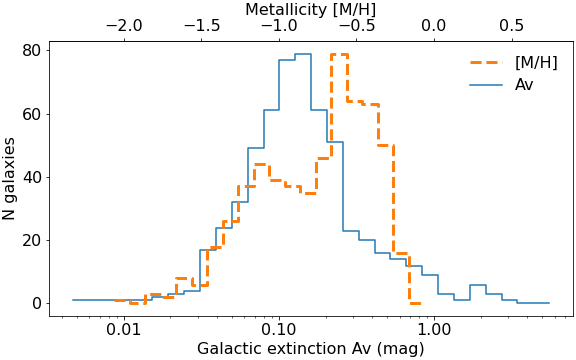}
    \caption{Distribution of some properties of the sample of 575 HST-observed galaxies. (Top) Semi-major-axis $a$ and distance from HECATE. (Bottom) Metallicity and Galactic extinction (see text for details).}
    \label{fig:gal_prop}
\end{figure}

To get a complete sample of nearby galaxies, we use the HECATE catalogue of galaxies (Heraklion Extragalactic CATaloguE, \citealt{Kovlakas2021}), based on the HyperLEDA database \citep{Paturel2003} but supplemented with robust distance estimates, galaxy mass, metallicity and star formation rate. The base sample, containing 1883 galaxies closer than 20~Mpc, is crossmatched with HST observations available on the \textit{Mikulski Archive for Space Telescopes} (MAST), to find galaxies with overlapping deep ($t_\mathrm{exp}>300$~s) exposures in two optical filters\footnote{This crossmatch was performed using the API of MAST, available as part of the \texttt{astroquery} v0.4.6 package \citep{astroquery}.}. Only galaxies having at least 2\% of their area covered by HST are considered, discarding \textit{de facto} the Magellanic Clouds\footnote{Recent studies targeting the Magellanic Clouds using ground-based surveys and \textit{Swift}-UVOT provide YSG samples of good completeness \citep{Neugent2010,Neugent2012,Humphreys2023,OGrady2024}.}. The list of selected filters is detailed in Table \ref{tab:filt}, together with the number of galaxies selected for each filter. We select the $F475W$, $F555W$, $F606W$ and $F814W$ filters because they have responses close to the Johnson-Cousins $V$ ($F475W$, $F555W$, $F606W$) and $I$ ($F814W$) filters, allowing us to study $V$ against $V-I$ colour-magnitude diagrams (CMD). The resulting HST-observed sample contains 575 galaxies. The distributions of distance and angular sizes are shown in the top panel of Figure \ref{fig:gal_prop}.

\subsection{HST data retrieval}

\begin{figure}
    \centering
    \includegraphics[width=8.5cm]{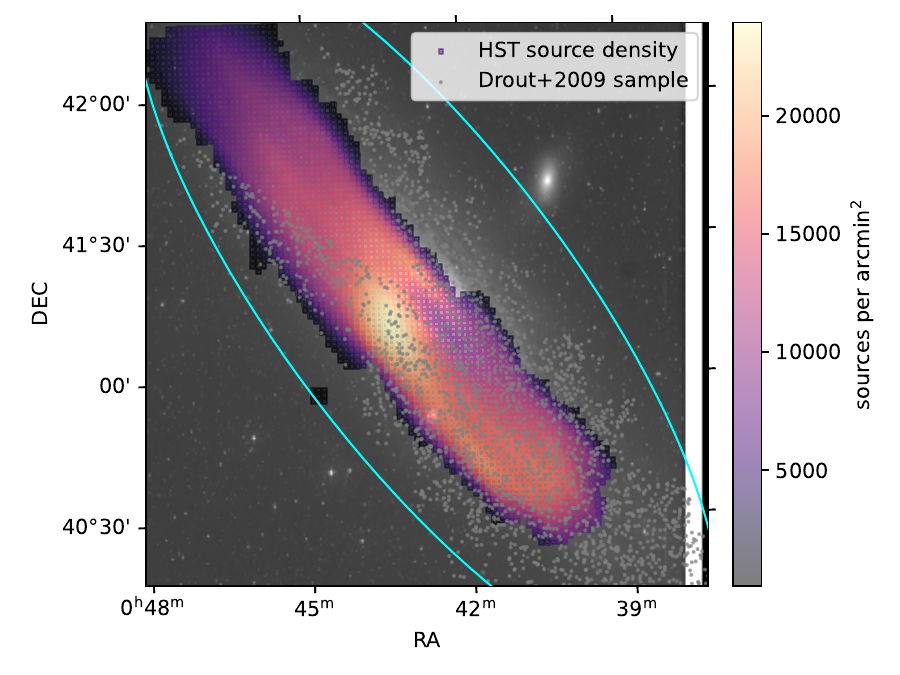}
    \caption{Density of HST sources under study for the galaxy M31, as retrieved through HSCv3 and MAST databases. The background image is from the Digitized Sky Survey 2 (DSS2, \citealt{Lasker1996}).}
    \label{fig:skyplot_m31}
\end{figure}

To retrieve HST sources, we use two databases comprising third-party catalogues in which source extraction has already been performed. The third release of the Hubble Source Catalog (HSCv3, \citealt{Whitmore2016}) includes \textit{Advanced Camera for Surveys} (ACS) and \textit{Wide-Field Camera 3} (WFC3) data that were public as of 2017 October 1. As such, it contains data for 449 of our galaxies observed before this date. HSCv3 data were retrieved using MAST queries with the CasJobs API\footnote{\url{https://github.com/rlwastro/mastcasjobs}}. For 118 additional galaxies, whose observation requirements are met only after 2017, we retrieve their source catalogues from MAST. At the time of the writing, eight galaxies could not be retrieved in these databases, due to proprietary exposures or missing source catalogues. As an example, the southern region of M31, which has sparse coverage in HSCv3, is included in the MAST retrieval. This leads to uniform coverage of the inner galaxy, as shown in Figure \ref{fig:skyplot_m31}). 

The MAST third party catalogues were produced at the time of the data reduction of the corresponding observations, using either DAOPHOT \citep{daophot}, Sextractor \citep{Sextractor}, or the Hubble Advanced Products pipeline \citep{HAP_pipeline}. For both types of catalogues, magnitudes are retrieved in the large aperture photometry (\textit{MagAper2}). We made sure that MAST catalogues are reliable and have well calibrated photometry, by comparing their sky coverage and their CMD with HSCv3 data for two test galaxies (NGC 45 and ESO 209-9). In all tested filters, the bias and standard deviation between HSCv3 and MAST photometry is $<0.05$~mag and $<0.2$~mag, respectively. Once downloaded, all catalogues of a given galaxy are merged and crossmatched to obtain a single ``master'' catalogue, using \texttt{astropy} v5.3.3 \citep{astropy}\footnote{A data-dependent matching radius $r_1$ ($0.3<\frac{r_1}{\mathrm{arcsec}}<2$) is used to associate sources from different filters, and another one $r_2$ ($0.1<\frac{r_2}{\mathrm{arcsec}}<0.3$) is used to eliminate duplicate sources.  $r_2$ is the local minimum of the distribution of separations in the interval [0.1,0.3] arcsec. This choice allows us to eliminate duplicate sources while keeping close neighbors, given the different levels of astrometric accuracies when using different source-extracting algorithms (sextractor, daophot).}. To discard compact star clusters, cosmic rays and sources affected by confusion, we discarded sources with concentration indexes (CI) $<0.85$ or $>1.5$ or magnitude errors $>0.1$.  

\begin{table}
    \caption{\label{tab:filt}Summary of HST Filters and Galaxy Selection.}
    \centering
    \begin{tabular}{ccccl}
    \hline\hline\\ [-1.5 ex]
         Instrument&  Red filter & Green Filter &N galaxies\\\hline\\ [-1.5 ex]
         ACS&  $F814W$& $F606W$ &210\\
 & &$F555W$ &40\\
 & &$F475W$ &47\\\hline\\ [-1.5 ex]
 WFC3& $F814W$&$F606W$&14\\
 & &$F555W$&54\\
 & &$F475W$ &23\\\hline\\ [-1.5 ex]
 WFPC2& $F814W$&$F606W$&116\\
 & &$F555W$ &71\\\hline
\end{tabular}
\tablefoot{HST filters used in this study. The last column gives the number of galaxies selected in the corresponding filters.}
\end{table}

\subsection{CMD analysis}
\label{sec:cmd_analysis}
\subsubsection{Stellar evolution tracks}

To analyse the CMD of galaxies, a physical reference is needed, such as isochrones or stellar evolution tracks in the observer's frame. However, the shape of an observed CMD heavily depends on intrinsic and extrinsic parameters. Intrinsic parameters include the galaxy star formation history and its metallicity, while extrinsic parameters correspond to measurement uncertainties and the extinction along the line of sight, commonly modelled with the extinction parameter $A_\lambda$. Using the synthetic photometry of stellar evolution tracks provided by MIST (MESA Isochrones \& Stellar Tracks\footnote{\url{https://waps.cfa.harvard.edu/MIST}}, \citealt{mist0,mist1}),
we generated tracks for a grid of zero-age main sequence (ZAMS) stellar masses, extinctions, and metallicities, summarised in Table \ref{tab:mist_param}. The ranges of extinctions and metallicities correspond to the ranges of estimated Galactic extinctions and metallicities of galaxies in our sample, and the mass range encompasses most of known LRNe progenitors (e.g. \citealt{MacLeod2022}). The upper limit of the mass range does not restrict the YSG selection; instead, it marks the point beyond which stellar masses become poorly estimated, as the Hertzsprung gap can no longer be consistently defined as a continuous post-MS phase. It is important to note that our selection will contain YSG regardless of their stellar evolution phase, either post-MS or post-RSG. While some YSG have been identified as post-RSG in recent studies (e.g. \citealt{Humphreys2023}), we cannot distinguish between them in our catalog. However, post-RSG YSG typically show only a fast, low-amplitude variability, therefore we expect a stable behaviour in future LSST observations.

\begin{table}
    \centering
    \caption{\label{tab:mist_param}MIST parameter grid.}
    \begin{tabular}{cc}
    \hline\hline\\ [-1.5 ex]
         Parameter&  Values\\\hline\\ [-1.5 ex]
         ZAMS mass M$_*$/M$_\odot$ & 
    $\{3, 4, 6, 8, 10, 12, 14, 16, 18, 20\}$\\
 Metallicity $Z$&$\{-1.2, -0.9, -0.6, -0.3, 0\}$\\
 Extinction $A_V$&$\{0, 0.2, 0.4, 0.6, 0.8, 1, 1.5, 2\}$\\\hline
 \end{tabular}

\end{table}

To estimate extinctions along the line-of-sight for each HECATE galaxy, we use the 2D dust map of \cite{Delchambre2022} released as part of \textit{Gaia} DR3. Metallicities and galaxy masses are provided in HECATE for only 15\% (276) and 62\% (1159) of our base sample, respectively.  To estimate metallicities, we therefore relied on the absolute $B$ magnitude -- oxygen abundance relations of \cite{Pilyugin2004} (Equations 12 and 15 for spiral and irregular galaxies, respectively\footnote{Given that only 12 galaxies in our sample are elliptical, we apply this relation accordingly}): $12+[O/H]=\min(5.8-0.139~B_\mathrm{abs}, 6.93-0.079~B_\mathrm{abs})$. Gas-phase oxygen abundances were converted to metallicity using the relation $Z=[M/H]=12+[O/H]-8.69$, where 8.69 is the solar oxygen abundance given in \cite{Asplund2009}. The resulting distributions of Galactic extinction and metallicity are shown in Figure \ref{fig:gal_prop} (bottom panel). In the following, all CMDs are in the Vega magnitude system, aperture-corrected (using the table\footnote{\url{https://archive.stsci.edu/hst/hsc/help/HSC\_Faq/ci_ap_cor_table_2016.txt}} recommended by the HSCv3 documentation) and de-reddened of Galactic extinction.

\subsubsection{HR gap definition}
\label{sec:gapdef}

\begin{figure}
    \centering
    \includegraphics[width=8.5cm]{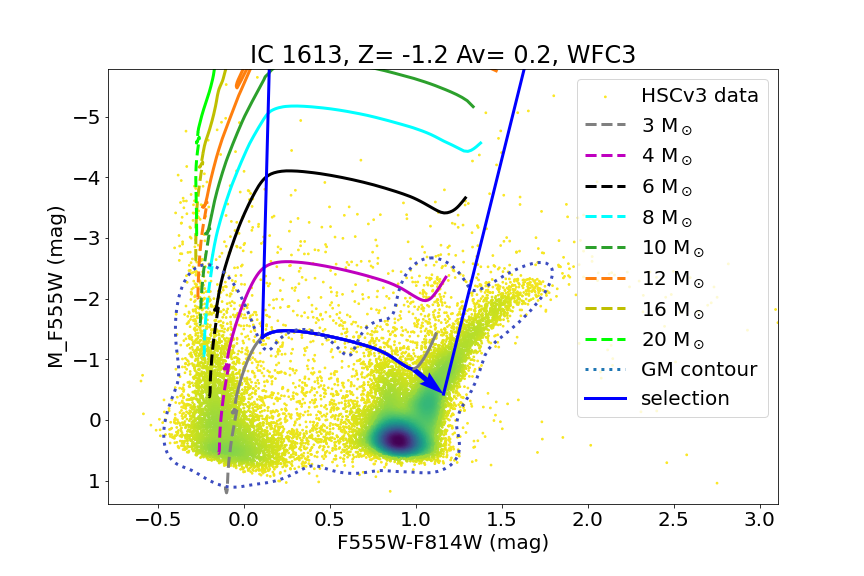}
    \caption{An example of CMD analysis. Extinction-corrected CMD of IC 1613 sources with MIST tracks for a metalllicity Z=-1.2 and an extinction of $A_V$=0.2. Hertzsprung gap candidates are selected above blue lines. The dotted contour shows the Gaussian mixture model representing the data.}
    \label{fig:cmdIC1613}
\end{figure}

The Hertzsprung gap of each galaxy is identified using the MIST tracks at different masses. Specifically, we define this gap as the time interval between two points, the post-MS bright turnoff (the point when the $V$-band luminosity first reaches 95\% of its post-MS bright turnoff value), and the pre-RGB faint turnoff, as illustrated in Figure \ref{fig:cmdIC1613} by the left-hand and right-hand solid blue lines. To account for the internal extinction of the galaxy and retrieve extinguished YSG candidates, we artificially applied a 0.4 mag extinction to the red-side selection cut, using a relative visibility value of $R(V) = 3.1$. Although this additional extinction is arbitrary, it ensures the retrieval of most extinguished candidates, as discussed in Section \ref{sec:contam}. YSG candidates are searched 
between these lines and above the MIST track of a reference ZAMS star mass, which ranges from 3 to 16 M$_\odot$ depending on the galaxy. To ensure the completeness of the selected samples at this reference mass, it is chosen to be the smallest mass for which the faint end of the Hertzsprung gap is well-observed, i.e., significantly brighter than the sensitivity of HST exposures. Consequently, we exclude 102 galaxies where the available HST data are too shallow to accurately detect YSGs with masses of 16 $M_\odot$ or below. 

\subsubsection{Statistical modelling}

Due to the significant uncertainties in the estimate of distance, metallicity, Galactic extinction, HST photometry of faint sources, and stellar evolution models, a mismatch between the true locus of the observed Hertzsprung gap and its MIST estimate is inevitable for some galaxies. In particular, some galaxies have part of their MS and/or RGB intersecting the selected area of the CMD. To exclude these densely populated regions, we adopt a similar approach as the YSG selection method of \cite{Addison2022}. The CMD of each galaxy is statistically modelled as a mixture of gaussian components (Gaussian Mixture, GM). The number of components is determined based on the logarithm of the number of sources, balancing the need to avoid overfitting in sparse CMDs while allowing for the fitting of complex structures in dense CMDs. To better retrieve the tails of the distribution (i.e. CMD features such as the RGB), the faintest and most populated regions of dense CMDs, as estimated by Kernel Density Estimation (KDE), were artificially depopulated by a factor of up to 20.
This step only removes sources in faint regions and not the YSG selection area.

To exclude the dense regions of the CMD, sources located within the GM contour of a reference likelihood are discarded. This contour intersects the Hertzsprung gap of the reference MIST track, with the likelihood value optimised through visual inspection. The resulting contour for IC 1613 is shown by the dotted line in Figure \ref{fig:cmdIC1613}, where the reference MIST track corresponds to a $3 \mathrm{M}_\odot$ star. The red giant branch is well excluded from the YSG selection area. A representative subset of the CMDs analysed in this study is shown in Appendix \ref{sec:appendix}. This statistical modelling requires a minimum of 100 sources to provide a reliable CMD match. Therefore, we discarded 96 galaxies, leaving 369 galaxies in our sample.

\subsection{Sample cleaning}
\label{sec:contam}
\subsubsection{Removing known contaminants}

Different types of contaminants may be included at this stage of the selection. First, the line of sight chance alignment of Milky Way stars, notably yellow dwarfs and white dwarfs, with the galaxy under consideration can cause them to appear within the Hertzsprung gap of its CMD. Likewise, background objects such as quasars can also appear in this locus. Second, the extinction affecting main-sequence stars inside each galaxy can displace them from the MS to the Hertzsprung gap in the observed CMD. Additionally, objects other than YSG also naturally appear in the Hertzsprung gap, such as cepheids and other variables within the instability strip, luminous blue variables (LBV), or unresolved globular clusters \citep[e.g.][]{Kraft1966, Justham2014, Mora2007}.

As an initial cleaning process, we crossmatched between our sample and the \textit{Simbad}\footnote{\url{https://simbad.cds.unistra.fr/}} and \textit{Gaia DR3} \citep{GaiaCollab2023} catalogues using a matching radius of 0.5''. \textit{Simbad} quasars, globular clusters, and classical cepheids were discarded. Additional quasars were found and excluded using the \textit{Milliquas} \citep{Flesch2021} and \textit{Gaia} Extragalactic \citep{GaiaExtragal2023} catalogues. However, we flagged known LBVs, YSGs and red supergiants (RSG) for further consideration within the sample. Sources exhibiting significant \textit{Gaia} proper motion ($\mathrm{PM}/\mathrm{PM_{err}}>4$) are identified as Galactic and discarded. 
Although other types of contaminants cannot be directly excluded, their presence can be quantified statistically.

\subsubsection{Quantifying foreground contamination}

To estimate the line of sight foreground contamination towards each galaxy, we utilised the TRILEGAL (TRIdimensional modeL of thE GALaxy, \citealt{Girardi2005}) simulation, which models the Milky Way's stellar content, incorporating the Sun's position and various Galactic components. It includes the thin disc, thick disc, halo, and bulge, along with their star formation histories. This simulation, notably used  by \cite{DalTio2022} to simulate the Milky Way as observed by the upcoming LSST survey, and accessible through an API from NOIRLab\footnote{\url{https://datalab.noirlab.edu/lsst_sim/index.php}}, simulates the photometry and stellar parameters of the observed Galactic population along designated lines of sight. Patches of 0.25 deg$^{2}$ were queried for each galaxy in our sample. Given that LSST's coverage is mostly restricted to the Southern Hemisphere (Dec<0), some galaxies lie outside the simulated regions. For cases where LSST covers the symmetric position relative to the Sun-Galactic poles plane, we used this position instead, assuming similar results due to the symmetric nature of TRILEGAL components. This planar symmetry was verified using test coordinates through the TRILEGAL web interface\footnote{Available at \url{http://stev.oapd.inaf.it/cgi-bin/trilegal}}. For the 30\% of galaxies (106 in total) still uncovered, we utilised the TRILEGAL web interface with the same parameters as \cite{DalTio2022}. Notably, the halo profile in the web interface differs from the LSST simulation, employing a $r^{-1/4}$ profile versus \citeauthor{DalTio2022}'s $r^{-1/2.5}$ profile. We adjusted for this by approximating the profile with $r^{-1/4}$, $\Omega$=0.0025~$M_\odot$~pc$^{-3}$ and $r_h$=5700 pc, based on comparable star counts and magnitude distributions.

\begin{figure}
    \centering
    \includegraphics[width=8.5cm]{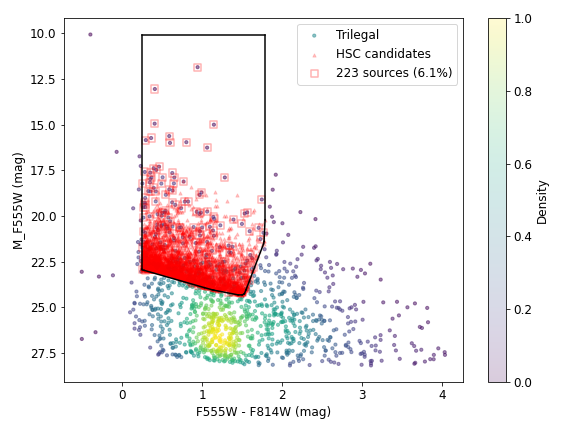}
    \caption{Illustration of the estimation of foreground contamination rate using the TRILEGAL Milky Way simulations. The CMD shows TRILEGAL sources overlapping with the footprint of M83 HSC sources. Highlighted in red are the YSG candidates identified within this sample. TRILEGAL sources falling within the black polygon, circled in red, are identified as foreground contaminants. They represent 3.1\% of the M83 YSG candidates.}
    \label{fig:cmdforeground}
\end{figure}

TRILEGAL sources matching the footprint of the selected HST exposures and exhibiting synthetic HST photometry within the specific Hertzsprung gap selection region are identified as contaminants. Figure \ref{fig:cmdforeground} illustrates this method for M83, showing the CMD of TRILEGAL sources overlapping with HST exposures and overlaying the observed YSG candidates. The results of this foreground contamination analysis are presented in Section \ref{sec:contamres}. 

\subsubsection{Extinguished main-sequence stars}

\begin{figure}
    \centering
    \includegraphics[width=8.5cm]{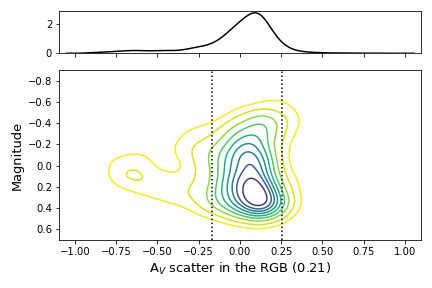}
    \includegraphics[width=8.5cm]{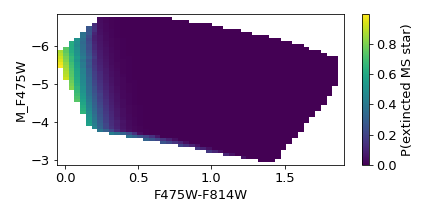}
    \caption{(Top) Density plot of the RGB of IC 1613 in the (Magnitude - relative $A_V$) plane. The x-axis zeropoint corresponds to the mean of the distribution. The top curve shows the KDE of $A_V$ values, with a standard deviation of $\pm$0.21 (dotted lines). (Bottom) Resulting probability map to be an extinguished MS star in the selection region of YSG sources in the IC 1613 galaxy.}
    \label{fig:ebv_scatter}
\end{figure}

\begin{figure}
    \centering
    \includegraphics[width=8.5cm]{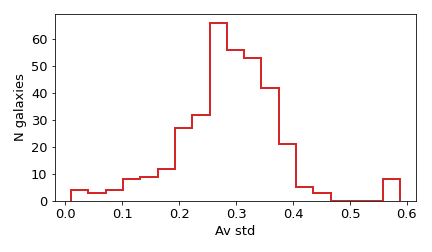}
    \caption{Distribution of the standard deviations of extinction for our galaxy sample.}
    \label{fig:histebv}
\end{figure}

Although we account for the Galactic extinction along the line of sight for each galaxy using the appropriate MIST model, intrinsic extinction inside each galaxy is not modelled at this stage. Consequently, a significant fraction of Hertzsprung gap sources may actually be extinguished MS stars. To quantify this, we assigned to each candidate a probability to be an extinguished main-sequence star. Assuming a normal distribution for extinction, we measured extinction scatter by evaluating the standard deviation of 
$A_V$  in a thin region of the CMD, chosen to be the red giant branch, as illustrated in Figure \ref{fig:ebv_scatter}. This standard deviation is used to model the additional extinction in the selection region, resulting in a probability map for each CMD (Figure \ref{fig:ebv_scatter}, bottom panel). This probability is naturally highest at the blue edge of the selection region, which separates the main-sequence region from the Hertzsprung gap. By averaging these probabilities on a per-galaxy basis, we can determine the contamination rate due to extinction. The distribution of $A_V$ scatters estimated for all selected galaxies is shown in Figure \ref{fig:histebv}. Besides providing contamination probabilities, it also suggests that the 0.4 mag extinction applied to the red edge of the selection region is sufficient to recover most of the extinguished YSG (Section \ref{sec:gapdef}). As a result, some RSGs are included in the sample, as discussed in Section \ref{sec:spectroYSG} for the case of M31. However, obscured YSGs and RSGs have also been identified as progenitors of certain infrared transients thought to share similarities with LRNe \citep{Jencson2019}.


\section{Results}

\label{sec:results}
\subsection{Selected candidates}



\begin{figure}
    \centering
    \includegraphics[width=8.5cm]{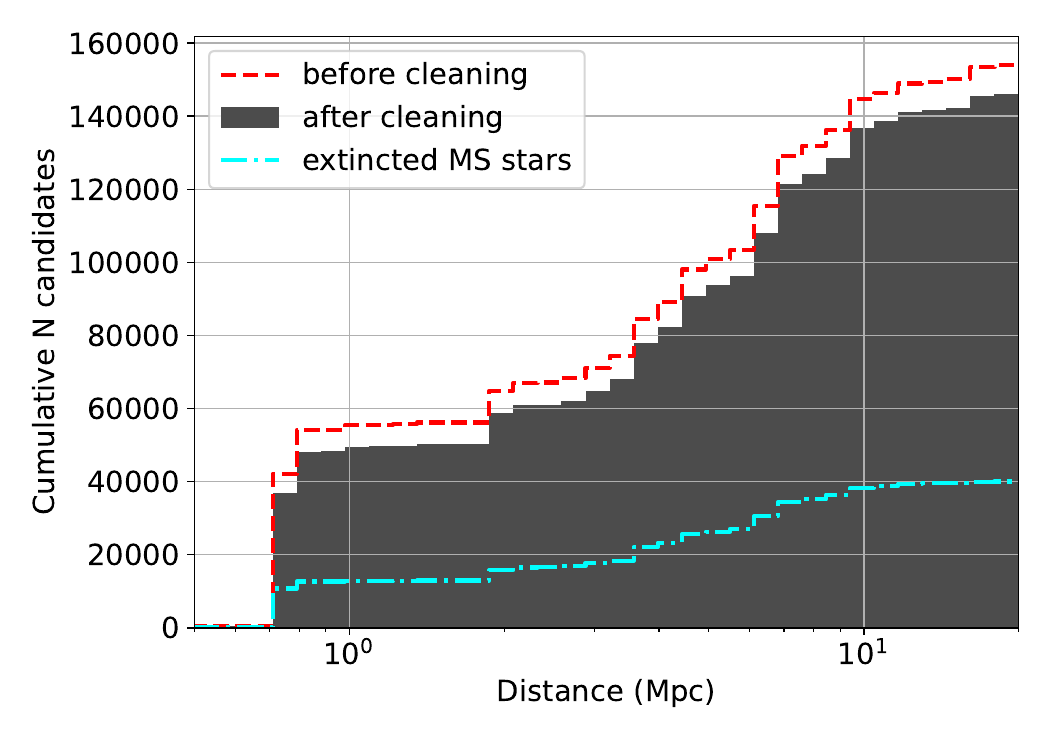}
    \caption{Cumulative distribution of distances for YSG candidates before and after cleaning, and estimated number of extinguished main-sequence stars.}
    \label{fig:cumulativedist}
\end{figure}

Our analysis identified a total of 154,494 YSG candidates before cleaning, a portion of which being listed in Table \ref{tab:candlist}. The cumulative distribution of their distances is depicted in Figure \ref{fig:cumulativedist}, indicating that about 25\% of these candidates are located in galaxies at approximately 0.8 Mpc. These candidates are actually located in M31 (40,566 candidates) and M33 (11,603), the two galaxies with most candidates. The three next galaxies with the most YSG candidates are M101 (8,885), NGC 300 (7,525), and NGC 253 (4,000).

As a result of the CMD analysis, we identified YSG candidates in 353 galaxies. The HST data provenance, available blue/green filter, coverage, and the parameters used for fitting the CMD as described in Section \ref{sec:cmd_analysis} are provided for each galaxy in Table \ref{tab:gallist}. This table also details the number of candidates for each galaxy and the estimated contribution from different contaminants.

\begin{table}
\caption{\label{tab:candlist}List of YSG candidates.}
    \centering
    \resizebox{\columnwidth}{!}{
    \begin{tabular}{cccccccc}
    \hline\hline\\ [-1.5 ex]
    ID & Galaxy & ra & dec & $m_{F814W}$ & M$_*$ & log(T$_\mathrm{eff}$ (K)) & P$_\mathrm{extin}$ \\
    & & (deg) & (deg) & (mag) & ($M_\odot$) & (K) \\\hline\\ [-1.5 ex]
1 & WLM Galaxy & 0.46607 & -15.50511 & 20.7 & 6.4 & 3.70 & 0.000\\
2 & WLM Galaxy & 0.46950 & -15.52687 & 21.1 & 4.9 & 3.82 & 0.001\\
3 & WLM Galaxy & 0.47109 & -15.50277 & 21.0 & 5.0 & 3.77 & 0.000\\
4 & WLM Galaxy & 0.47403 & -15.52111 & 21.3 & 5.1 & 3.71 & 0.000\\
5 & WLM Galaxy & 0.47516 & -15.53001 & 21.3 & 4.7 & 3.83 & 0.004\\
6 & WLM Galaxy & 0.47576 & -15.50463 & 21.8 & 4.3 & 3.92 & 0.722\\
7 & WLM Galaxy & 0.47651 & -15.50017 & 20.2 & 6.0 & 3.74 & 0.000\\
8 & WLM Galaxy & 0.47771 & -15.51067 & 21.2 & 4.7 & 3.81 & 0.000\\
9 & WLM Galaxy & 0.47795 & -15.49877 & 20.5 & 5.6 & 3.73 & 0.000\\
10 & WLM Galaxy & 0.47857 & -15.52563 & 22.0 & 4.1 & 3.93 & 0.881\\
11 & WLM Galaxy & 0.48087 & -15.49872 & 21.6 & 4.5 & 3.90 & 0.463\\
12 & WLM Galaxy & 0.48175 & -15.49939 & 19.9 & 6.7 & 3.84 & 0.023\\
13 & WLM Galaxy & 0.48180 & -15.50023 & 20.3 & 5.9 & 3.79 & 0.000\\
14 & WLM Galaxy & 0.48257 & -15.50483 & 21.2 & 5.1 & 3.90 & 0.430\\
15 & WLM Galaxy & 0.48324 & -15.51946 & 18.9 & 8.4 & 3.72 & 0.000\\
16 & WLM Galaxy & 0.48382 & -15.50083 & 20.8 & 6.0 & 3.70 & 0.000\\
17 & WLM Galaxy & 0.48422 & -15.52093 & 21.0 & 5.1 & 3.84 & 0.016\\
18 & WLM Galaxy & 0.48458 & -15.50045 & 20.7 & 6.0 & 3.70 & 0.000\\
19 & WLM Galaxy & 0.48469 & -15.50093 & 20.4 & 5.7 & 3.79 & 0.000\\
20 & WLM Galaxy & 0.48472 & -15.53525 & 20.9 & 5.1 & 3.77 & 0.000\\
21 & WLM Galaxy & 0.48475 & -15.49602 & 20.4 & 5.8 & 3.82 & 0.002\\
22 & WLM Galaxy & 0.48613 & -15.51076 & 21.3 & 4.5 & 3.74 & 0.000\\
23 & WLM Galaxy & 0.48639 & -15.50061 & 21.2 & 5.4 & 3.71 & 0.000\\
24 & WLM Galaxy & 0.48646 & -15.52779 & 21.2 & 5.4 & 3.70 & 0.000\\
25 & WLM Galaxy & 0.48668 & -15.49562 & 21.3 & 5.1 & 3.70 & 0.000\\
\hline
    \end{tabular}
    }
    \tablefoot{Column 2: host galaxy. Columns 3--5: HST coordinates (deg) and magnitude (mag). Columns 6--7: MIST inferred stellar parameters. Column 8: Probability to be an extinguished MS star. The complete list will be made available in electronic form.}
    
\end{table}

\begin{table*}
\caption{\label{tab:gallist}Galaxies having more than 400 YSG candidates, sorted by distance.}
\resizebox{0.9\textwidth}{!}{
\begin{tabular}{ccccccccccccc}
\hline\hline\\ [-1.5 ex]
     Galaxy & Distance & HSTcov & Database & Camera / Filter & N$_\mathrm{src}$ & MIST M$_{*,\mathrm{ref}}$& $A_V$ & [Fe/H] & $n_\mathrm{GM}$ &  N$_\mathrm{cand}$ & $f_\mathrm{extin}$& $f_\mathrm{foreground}$\\
     & (Mpc) & (\%) & & & & ($M_\odot$) & (mag) & & & & & \\\hline\\ [-1.5 ex]
IC 10 & 0.7 & 49.1 & HSCv3 & ACS $F606W$ & 208629 & 6 & 1.42 & -1.30 & 10 & 485 & 0.214 & 0.237\\
M31 & 0.8 & 36.1 & HSCv3+MAST & ACS $F475W$ & 31697166 & 4 & 0.49 & -0.88 & 0 & 40566 & 0.260 & 0.128\\
M33 & 0.8 & 17.0 & HSCv3 & ACS $F475W$ & 2834135 & 4 & 0.51 & -1.13 & 10 & 11603 & 0.161 & 0.137\\
PegDIG & 1.0 & 83.5 & HSCv3 & ACS $F475W$ & 53724 & 3 & 0.28 & -1.39 & 90 & 804 & 0.033 & 0.010\\
NGC 300 & 1.9 & 63.0 & HSCv3 & ACS $F606W$ & 715947 & 4 & 0.05 & -0.62 & 50 & 7525 & 0.316 & 0.013\\
NGC 1569 & 1.9 & 92.0 & HSCv3 & ACS $F606W$ & 53637 & 6 & 0.81 & -0.89 & 50 & 1019 & 0.437 & 0.005\\
NGC 55 & 2.2 & 15.0 & HSCv3 & ACS $F606W$ & 146004 & 6 & 0.13 & -1.45 & 10 & 1897 & 0.366 & 0.003\\
UGC 9240 & 2.8 & 100.0 & HSCv3 & ACS $F606W$ & 25138 & 4 & 0.15 & -1.13 & 10 & 832 & 0.239 & 0.002\\
NGC 2403 & 3.2 & 34.2 & HSCv3 & ACS $F606W$ & 203205 & 8 & 0.27 & -0.48 & 50 & 1494 & 0.391 & 0.016\\
NGC 2366 & 3.3 & 97.6 & HSCv3 & ACS $F555W$ & 101143 & 6 & 0.20 & -1.14 & 10 & 650 & 0.139 & 0.013\\
UGC 4305 & 3.4 & 58.7 & HSCv3 & ACS $F555W$ & 97445 & 6 & -0.04 & -1.07 & 50 & 557 & 0.169 & 0.045\\
IC 342 & 3.4 & 6.9 & HSCv3 & ACS $F555W$ & 12201 & 8 & 1.38 & -0.87 & 0 & 449 & 0.152 & 0.202\\
NGC 7793 & 3.4 & 77.5 & HSCv3 & ACS $F555W$ & 90987 & 8 & 0.06 & -0.57 & 50 & 647 & 0.267 & 0.030\\
NGC 2976 & 3.8 & 67.0 & HSCv3 & ACS $F606W$ & 100684 & 8 & 0.37 & -0.71 & 50 & 571 & 0.460 & 0.003\\
M81 & 3.8 & 88.6 & HSCv3 & ACS $F606W$ & 958597 & 8 & 0.20 & -0.31 & 10 & 3357 & 0.344 & 0.043\\
NGC 253 & 3.9 & 38.8 & HSCv3 & ACS $F606W$ & 634872 & 8 & 0.37 & -0.93 & 10 & 4000 & 0.442 & 0.005\\
NGC 3077 & 3.9 & 55.4 & HSCv3 & ACS $F606W$ & 121976 & 8 & 0.37 & -0.67 & 50 & 495 & 0.297 & 0.006\\
NGC 4244 & 4.2 & 93.9 & HSCv3 & ACS $F606W$ & 152981 & 8 & 0.12 & -0.61 & 20 & 803 & 0.287 & 0.041\\
NGC 4449 & 4.3 & 84.4 & HSCv3 & ACS $F555W$ & 142096 & 8 & 0.14 & -0.79 & 20 & 2022 & 0.283 & 0.004\\
NGC 1313 & 4.3 & 69.7 & HSCv3 & ACS $F555W$ & 168305 & 8 & 0.19 & -0.80 & 0 & 1491 & 0.211 & 0.043\\
ESO 223-9 & 4.4 & 98.2 & HSCv3 & ACS $F606W$ & 15109 & 8 & 0.60 & -1.05 & 90 & 441 & 0.263 & 0.525\\
M94 & 4.6 & 34.3 & HSCv3 & ACS $F555W$ & 17481 & 8 & 0.07 & -0.35 & 50 & 496 & 0.300 & 0.035\\
ESO 137-18 & 4.7 & 91.9 & HSCv3 & ACS $F606W$ & 19885 & 8 & 0.54 & -0.83 & 30 & 441 & 0.176 & 0.452\\
M83 & 4.8 & 54.6 & HSCv3 & ACS $F555W$ & 201354 & 8 & 0.16 & -0.83 & 10 & 3671 & 0.349 & 0.031\\
NGC 4395 & 4.8 & 13.3 & HSCv3 & ACS $F606W$ & 61908 & 6 & 0.10 & -0.63 & 10 & 2604 & 0.146 & 0.009\\
NGC 4656 & 5.0 & 30.8 & HSCv3 & WFC3 $F555W$ & 18113 & 8 & -0.05 & -0.48 & 50 & 554 & 0.316 & 0.009\\
NGC 6946 & 5.5 & 56.2 & HSCv3 & WFC3 $F606W$ & 21289 & 8 & 0.97 & -1.34 & 0 & 551 & 0.462 & 0.075\\
NGC 6503 & 6.2 & 61.5 & HSCv3 & ACS $F606W$ & 39071 & 8 & -0.02 & -0.86 & 50 & 637 & 0.447 & 0.012\\
NGC 1566 & 6.5 & 41.6 & HSCv3 & ACS $F555W$ & 21654 & 8 & -0.01 & -0.63 & 20 & 1004 & 0.231 & 0.012\\
Fourcade-Figueroa & 6.7 & 43.3 & HSCv3 & ACS $F606W$ & 38012 & 8 & 0.13 & -1.26 & 0 & 446 & 0.308 & 0.137\\
M101 & 6.7 & 40.8 & HSCv3 & ACS $F555W$ & 374087 & 8 & 0.12 & -1.09 & 10 & 8885 & 0.278 & 0.018\\
IC 1727 & 6.9 & 59.4 & HSCv3 & ACS $F606W$ & 21623 & 8 & 0.18 & -0.97 & 40 & 608 & 0.220 & 0.022\\
NGC 3621 & 7.2 & 42.6 & HSCv3 & ACS $F555W$ & 67015 & 8 & 0.17 & -1.01 & 0 & 3089 & 0.398 & 0.011\\
M74 & 7.3 & 68.1 & HSCv3 & ACS $F555W$ & 59490 & 8 & 0.38 & -0.86 & 20 & 3265 & 0.267 & 0.007\\
IC 4710 & 7.4 & 97.8 & HSCv3 & ACS $F606W$ & 26877 & 8 & 0.18 & -0.73 & 20 & 569 & 0.280 & 0.127\\
M106 & 7.5 & 97.9 & HSCv3 & ACS $F555W$ & 66349 & 10 & 0.32 & -0.88 & 0 & 2173 & 0.244 & 0.029\\
NGC 4631 & 7.5 & 99.9 & HSCv3 & ACS $F606W$ & 192224 & 10 & 0.17 & -0.92 & 0 & 2496 & 0.269 & 0.003\\
NGC 2188 & 8.0 & 87.8 & HSCv3 & ACS $F606W$ & 16356 & 8 & 0.20 & -0.63 & 50 & 739 & 0.288 & 0.010\\
M51 & 8.6 & 50.9 & HSCv3 & ACS $F606W$ & 58919 & 12 & 0.64 & -0.46 & 0 & 576 & 0.295 & 0.016\\
NGC 4559 & 8.8 & 45.5 & HSCv3 & ACS $F606W$ & 47602 & 10 & 0.14 & -0.68 & 0 & 623 & 0.259 & 0.023\\
NGC 4485 & 8.9 & 99.5 & HSCv3 & ACS $F606W$ & 10838 & 8 & 0.10 & -0.65 & 10 & 456 & 0.340 & 0.004\\
NGC 2683 & 9.0 & 72.9 & HSCv3 & ACS $F606W$ & 22402 & 8 & -0.01 & -0.63 & 20 & 472 & 0.361 & 0.011\\
NGC 24 & 9.1 & 86.9 & HSCv3 & ACS $F606W$ & 18596 & 8 & 0.18 & -1.26 & 0 & 415 & 0.402 & 0.010\\
NGC 4490 & 9.6 & 33.9 & HSCv3 & WFC3 $F555W$ & 8312 & 8 & 0.15 & -0.59 & 20 & 2847 & 0.146 & 0.000\\
NGC 3344 & 9.8 & 39.2 & HSCv3 & ACS $F606W$ & 31474 & 10 & 0.37 & -0.98 & 0 & 1378 & 0.247 & 0.009\\
NGC 672 & 10.0 & 57.5 & HSCv3 & ACS $F606W$ & 32793 & 10 & 0.33 & -1.18 & 0 & 984 & 0.322 & 0.012\\
IC 5201 & 10.3 & 62.0 & HSCv3 & ACS $F606W$ & 22810 & 10 & 0.14 & -0.48 & 50 & 463 & 0.235 & 0.045\\
M66 & 10.4 & 70.4 & HSCv3 & WFC3 $F555W$ & 29283 & 12 & 0.15 & -0.56 & 0 & 705 & 0.170 & 0.004\\
NGC 5398 & 11.4 & 97.8 & HSCv3 & ACS $F606W$ & 14604 & 8 & 0.33 & -1.11 & 0 & 552 & 0.388 & 0.028\\
NGC 4051 & 12.2 & 48.2 & MAST & WFC3 $F555W$ & 5922 & 12 & 0.37 & -0.68 & 50 & 1007 & 0.198 & 0.015\\
NGC 4565 & 12.8 & 73.5 & HSCv3 & ACS $F606W$ & 42551 & 10 & 0.06 & -0.24 & 80 & 926 & 0.313 & 0.028\\

\hline
\end{tabular}
}
\tablefoot{Column 2: distance to the galaxy. Columns 3--6: information on their HST dataset. Columns 7--10: parameters used to fit their CMD. Column 11: number of resulting candidates. Columns 12--13: contamination estimates. The complete list of galaxies having YSG candidates will be made available in electronic form.}
    
\end{table*}

\begin{figure}
    \centering
    \includegraphics[width=8.5cm]{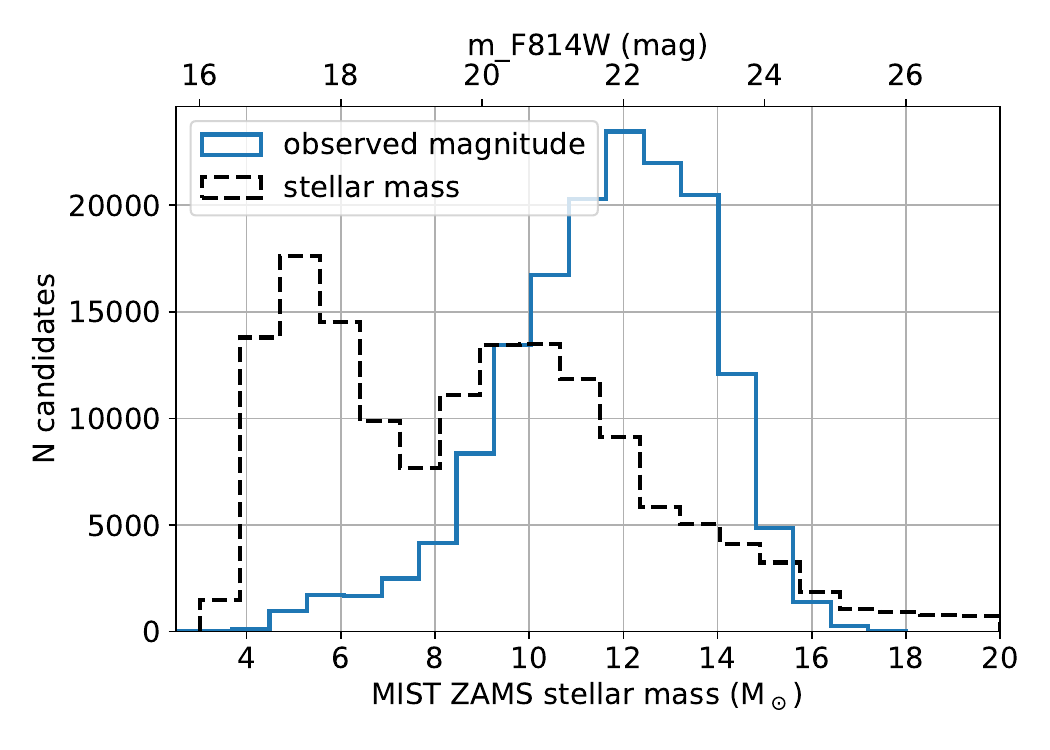}
    \caption{Distributions of observed $F814W$ magnitude and stellar masses of YSG candidates.}
    \label{fig:histmstarf814w}
\end{figure}

\begin{figure}
    \centering
    \includegraphics[width=8.5cm]{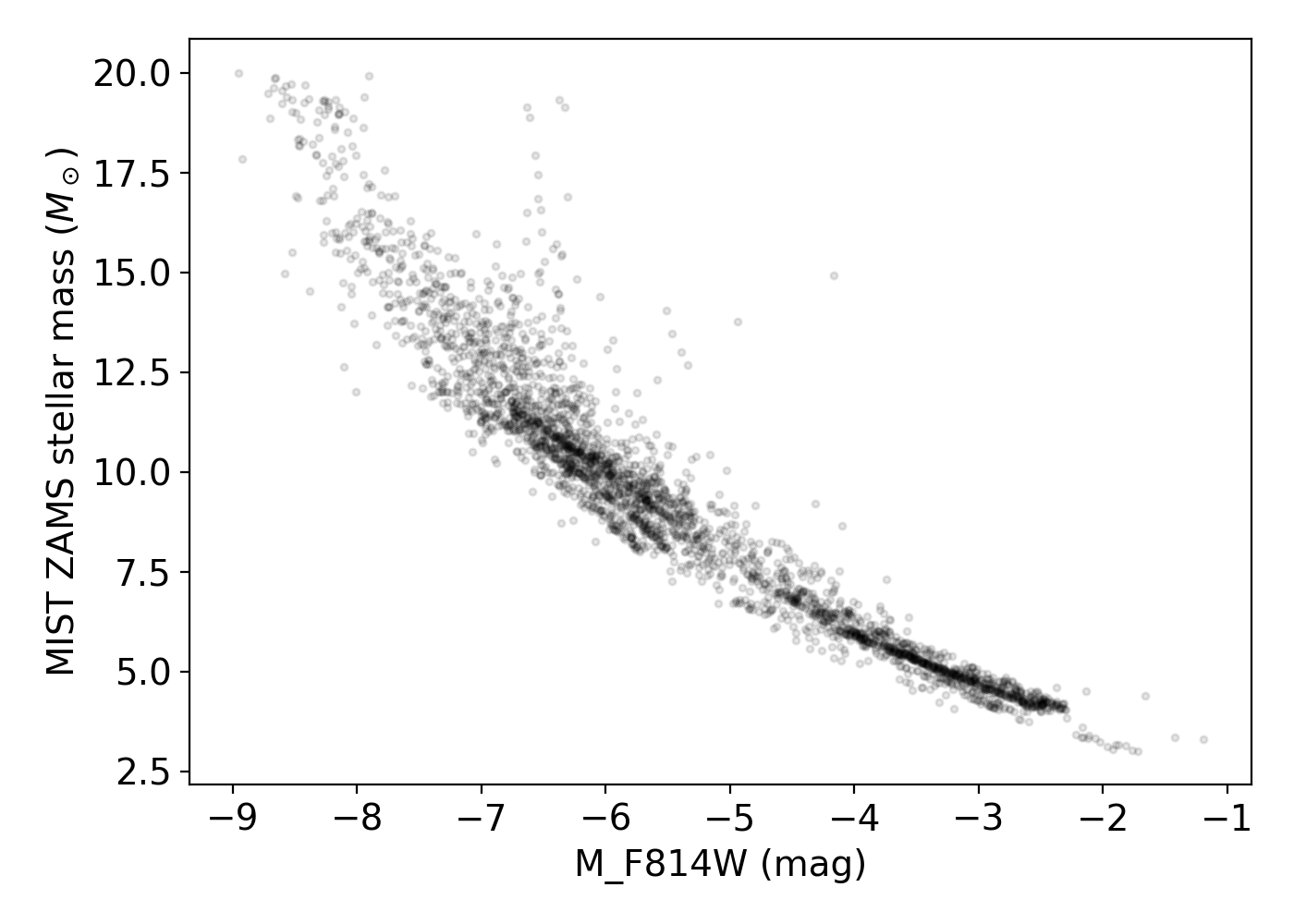}
    \caption{MIST inferred ZAMS stellar masses of YSG candidates as a function of their absolute $F814W$ magnitude.}
    \label{fig:relmstarf814w}
\end{figure}

YSG candidates exhibit apparent magnitudes in the $F814W$ band ranging from 16 to 25, with a peak at approximately 22.2 (Figure \ref{fig:histmstarf814w}). Single-star ZAMS stellar masses, luminosities, and temperatures for these candidates were estimated by interpolation using the MIST stellar evolutionary tracks within the Hertzsprung gap. The distribution of stellar masses, shown in Figure \ref{fig:histmstarf814w}, appears bimodal: a lower-mass peak, around approximately $5 M_\odot$, primarily corresponds to candidates in M31, M33 and NGC 253, and a higher-mass peak, around approximately $10 M_\odot$, corresponds to more distant candidates. This bimodality is further illustrated in Figure \ref{fig:relmstarf814w}, showing the tight correlation between interpolated stellar masses and $F814W$ absolute magnitude.

\subsection{Contamination fraction}
\label{sec:contamres}

\begin{figure}
    \centering
    \includegraphics[width=8.5cm]{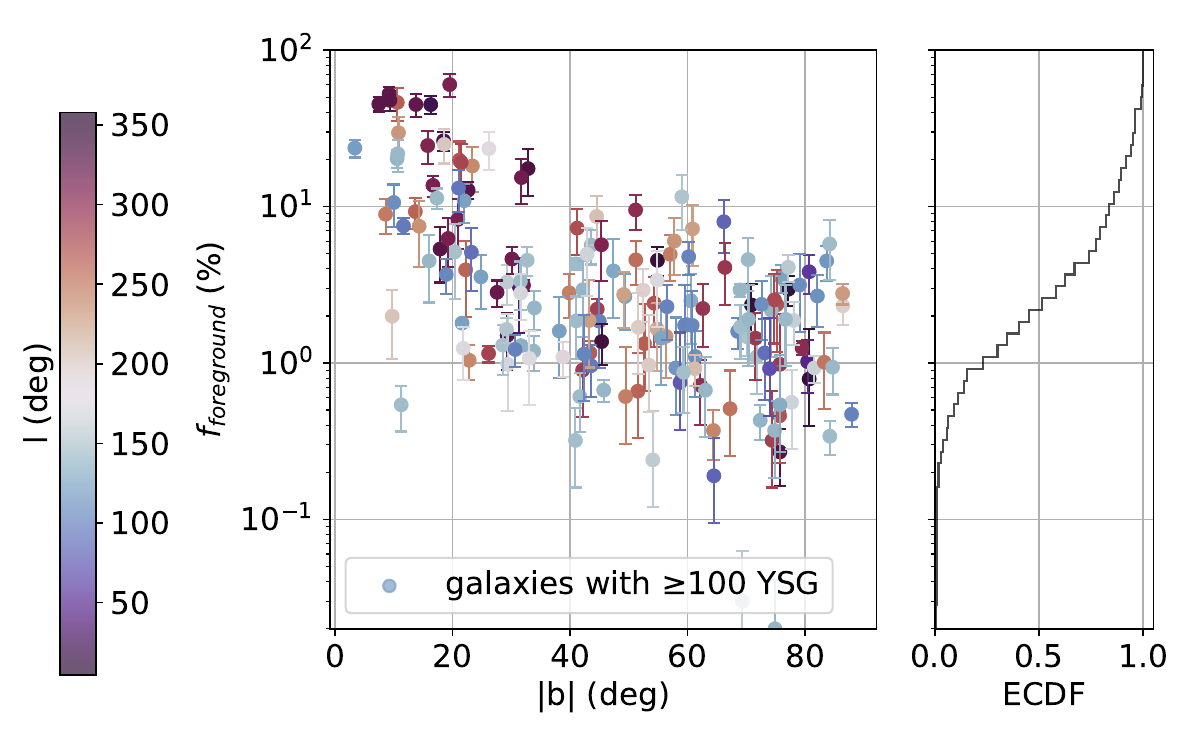}
    \caption{(Left) Foreground contamination rate as a function of Galactic latitude, using the TRILEGAL Milky Way simulations. The colour encodes the Galactic longitude. To maintain readability, only galaxies with a minimum of 100 YSG candidates are included. (Right) Empirical cumulative distribution function (ECDF) of the foreground contamination rate. 75\% of galaxies have less than 10\% of their YSG candidates being foreground contaminants.}
    \label{fig:fcontforeground}
\end{figure}

To evaluate the contamination fraction from intrinsic extinction and foreground sources, we considered various types of contaminants. Foreground contamination was quantified using the TRILEGAL star counts. According to the method detailed in Section \ref{sec:contam}, approximately 5\% of the sources in our YSG sample are expected to be foreground contaminants (including sources having high \textit{Gaia} proper motion). The foreground contamination mainly affects galaxies at low Galactic latitudes ($|b|<25$º) and 75\% of the galaxies have less than 10\% of their YSG candidates being foreground, 
as shown in Figure \ref{fig:fcontforeground} (right panel). As illustrated in this Figure, galaxies closer to the Galactic plane or toward the Galactic center exhibit higher contamination fractions. After using \textit{Gaia} proper motions to clean the sample, removing 6,623 sources (4.3\% of the sample), we therefore estimate the remaining foreground contamination to be $\lesssim$ 1\%. 

\label{sec:mscontam}

\begin{figure}
    \centering
    \includegraphics[width=8.5cm]{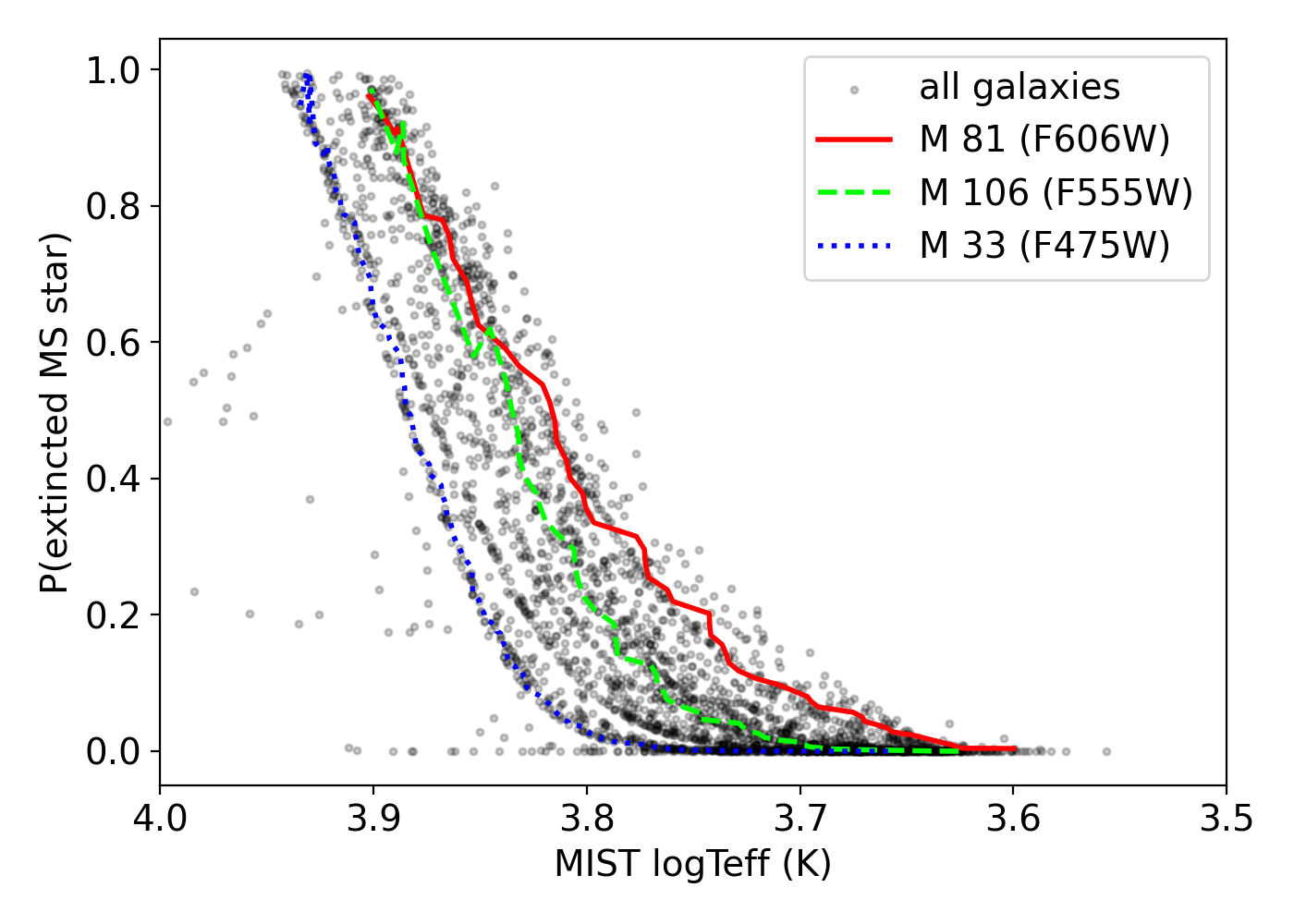}
    \caption{Probability of YSG candidates to be an extinguished MS star as a function of their MIST temperature. Three example galaxies are shown, illustrating the impact of the HST blue/green filter used to analyse the CMD.}
    \label{fig:PextinTeff}
\end{figure}

Additionally, the probability of each YSG candidate to be an extinguished main-sequence star was computed using the modelled intrinsic extinction (Section \ref{sec:contam}). The mean probability for all sources in our YSG catalogue is 0.25, with the fraction of contaminants remaining constant across all distances (Figure \ref{fig:cumulativedist}). As expected, this probability is significantly influenced by the temperature of the source, and the filter used in the CMD analysis, as depicted in Figure \ref{fig:PextinTeff}.

\begin{figure}
    \centering
    \includegraphics[width=8.5cm]{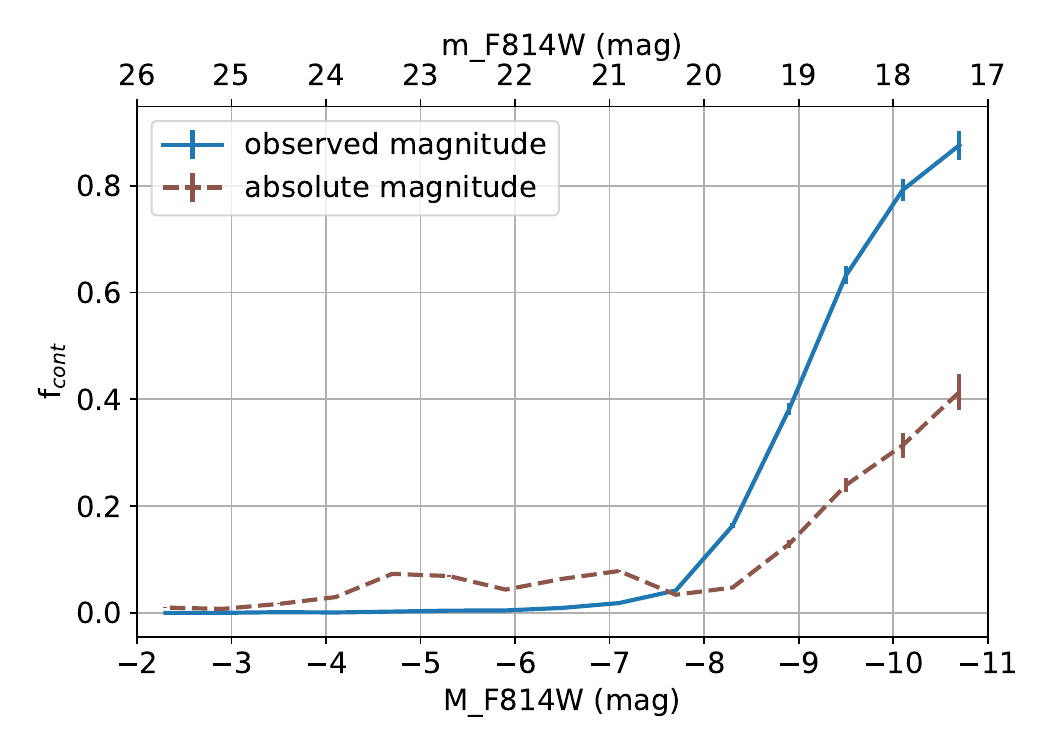}
    \caption{Fraction of identified contaminants using Simbad and \textit{Gaia} catalogues, as a function of observed and absolute $F814W$ magntiude.}
    \label{fig:fcontf814w}
\end{figure}

Finally, contaminants such as quasars, Cepheids and luminous blue variables were identified and removed using \textit{Gaia} and Simbad databases. These known contaminants represent a minority of the sample across all distances (Figure \ref{fig:cumulativedist}). These sources are typically brighter in both observed and absolute magnitudes, as shown in Figure \ref{fig:fcontf814w}. The clean sample of YSG contains 146,502 sources.

\section{Applications}
\label{sec:application}
\subsection{Completeness for known LRN progenitors}

\begin{table}
    \caption{\label{tab:lit_progenitors}Retrieval status of known LRN progenitors in the YSG sample.}
    \centering\footnotesize
    \begin{tabular}{ccll}
    \hline\hline\\ [-1.5 ex]
 LRN name& Distance (Mpc)&Status &Ref.\\\hline\\ [-1.5 ex]
         M31LRN2015&  0.77&retrieved
 &[1]\\
         AT2019zhd&0.78&not catalogued&[2]\\
         AT2020hat&  5.16&retrieved
 &[3]\\
         M101-2015OT&6.4&outside HST&[4]\\
         AT2018bwo&  6.8&retrieved
 &[5]\\
         AT2021biy&7.46&retrieved
 &[6]\\
         AT2021blu&  8.64&retrieved
 &[7]\\
         NGC4490-OT& 
    9.6&not catalogued&[8]\\\hline
    \end{tabular}
    \tablefoot{LRN progenitors identified in the literature, and their retrieval status in our YSG sample. References: [1] \citealt{MacLeod2017}, [2] \citealt{Pastorello2021}, [3] \citealt{Pastorello2021b}, [4] \citealt{Blagorodnova2017}, [5] \citealt{Blagorodnova2021}, [6] \citealt{Cai2022} [7] \citealt{Pastorello2023}, [8] \citealt{Smith2016}.}
\end{table}

Many extragalactic LRNe discovered during the last decade had pre-outburst images taken several years before the transient (e.g. \citealt{MacLeod2022} and references therein). For eight of them, listed in Table \ref{tab:lit_progenitors}, their host galaxy is in our sample, allowing us to assess the rate of retrieved progenitors in our sample of YSG candidates. Three out of eight progenitors are missing in our sample. For AT2015dl, the transient is located outside the footprint of the HST observations. For NGC4490-OT, although the progenitor was measured and studied by \cite{Smith2016MNRAS}, its apparent magnitude of $m_\mathrm{F606W}=23.58\pm$0.24 was too faint to be retrieved as an HSCv3 source (the faintest retrieved sources in a 30 arcsec radius circle around the progenitor position were 23.2 mag). For AT2019zhd, the progenitor had a low luminosity ($M_\mathrm{F555W}= 0.21 \pm 0.14$, \citealt{Pastorello2021}), placing it in the most populated region of the Hertzsprung gap. Consequently, it could not be selected after the Gaussian mixture cut. Moreover, its proximity to the edge of the field of view prevents this source from being retrieved as a MAST-catalogued source, and its magnitude error would either way exclude it from our sample. Considering these results, our sample appears to represent a relatively comprehensive selection of LRN progenitors, with a $\sim$70\% completeness level in regions covered by HST.

\subsection{Progenitors of past and ongoing transients: crossmatch to TNS}
\label{sec:tnsmatches}

\begin{table}
\caption{\label{tab:tnsxmatch}YSG candidates matching TNS objects.}
    \centering
    \resizebox{\columnwidth}{!}{\begin{tabular}{cccccccc}
    \hline\hline\\ [-1.5 ex]
 Galaxy & $m_{F814W}$ & TNS name & Sep. & TNS Filter & $m_\mathrm{TNS}$ & $M_\mathrm{TNS}$ & type\\ 
   & (mag) &  & (arcsec) &  & (mag) & (mag) & \\ \hline\\ [-1.5 ex]
NGC 3631 & 18.8 & 2016bau & 0.08 & Clear & 17.8 & -13.5 & SN Ib\\
NGC 2403 & 20.2 & 2016ccd & 0.16 & Clear & 18.0 & -9.5 & LBV?\\
M 31 & 21.4 & 2016fbx & 0.27 & Clear & 17.7 & -6.7 & Nova\\
NGC 45 & 22.6 & 2018bwo & 0.13 & Clear & 16.4 & -12.4 & LRN\\
NGC 4449 & 21.7 & 2018mmb & 0.55 & r & 19.7 & -8.4 & --\\
NGC 4449 & 20.0 & 2019ejn & 0.41 & Clear & 18.8 & -9.4 & LBV?\\
M 74 & 23.3 & 2019krl & 0.01 & g & 21.0 & -8.3 & --\\
M 31 & 20.0 & 2020aaqy & 0.35 & r & 20.9 & -3.5 & --\\
NGC 300 & 23.7 & 2020acli & 0.59 & Clear & 18.4 & -8.0 & Nova\\
M 31 & 18.7 & 2020adbp & 0.14 & r & 20.4 & -4.1 & --\\
NGC 5068 & 24.3 & 2020hat & 0.02 & orange & 17.8 & -10.7 & LRN\\
NGC 6503 & 21.6 & 2021ahsv & 0.17 & r & 20.1 & -8.8 & --\\
NGC 4631 & 21.4 & 2021biy & 0.13 & orange & 18.1 & -11.3 & LRN\\
Spider & 21.3 & 2021blu & 0.12 & orange & 18.5 & -11.0 & LRN\\
M 31 & 18.0 & 2021lzk & 0.24 & r & 19.3 & -5.2 & V*\\
M 31 & 19.8 & 2021tmm & 0.28 & r & 20.5 & -4.0 & --\\
M 31 & 19.6 & 2021ytf & 0.10 & r & 20.5 & -4.0 & --\\
NGC 253 & 19.5 & 2022llt & 0.06 & G & 19.0 & -9.0 & --\\
M 101 & 23.2 & 2023azz & 0.51 & g & 20.4 & -8.7 & --\\
M 31 & 21.7 & 2023wck & 0.52 & V & 21.2 & -3.2 & microlensing?\\
NGC 3310 & 19.4 & 2023wgz & 0.57 & Clear & 14.2 & -17.1 & --\\
M 31 & 21.8 & 2023wot & 0.58 & V & 20.4 & -4.0 & --\\
NGC 3621 & 22.4 & 2024ggi & 0.36 & orange & 18.9 & -10.4 & SN II\\
M 101 & 21.0 & 2024ikg & 0.30 & g & 20.5 & -8.6 & --\\

\hline
    \end{tabular}}
    \tablefoot{The last column shows the type according to TNS with eventual corrections (V*: variable star). Objects are sorted by TNS name.}
\end{table}

\begin{figure}
    \centering
    \includegraphics[width=8.5cm]{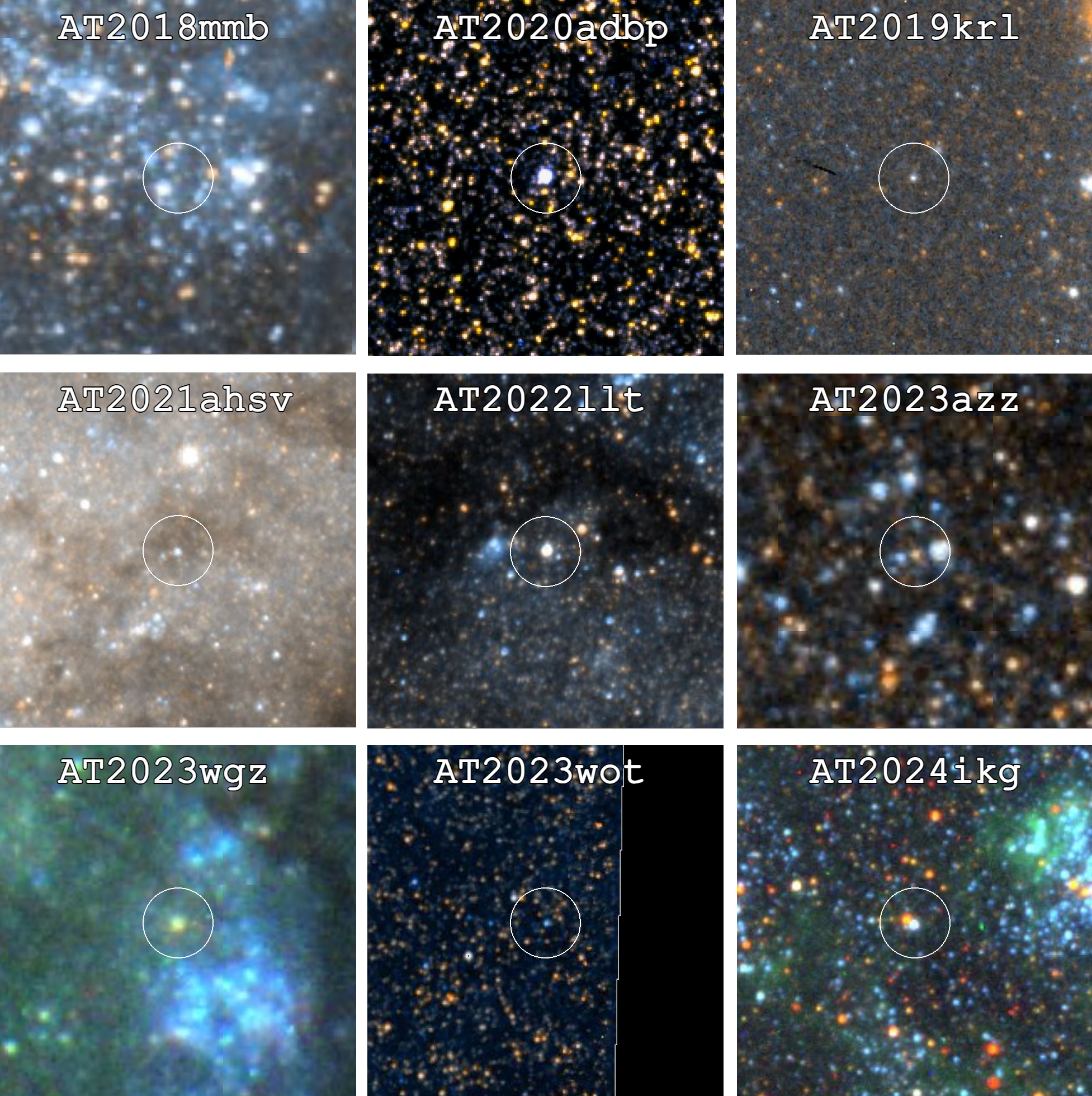}
    \caption{HST cutouts of the 12 candidates resulting from our TNS cross-match. Images are 10'' side-to-side, with a circle of 1'' radius pinpointing the location of the YSG source. Adapted from HLA colour composites.}
    \label{fig:hst_cutouts}
\end{figure}

Our YSG sample can be used to find possible progenitors for past transients and further analyse them. To this end, we crossmatched our YSG candidates to the TNS public objects as of 2024 August 31st, with a match radius of 0.6''. The resulting list of 24 transients is detailed in Table \ref{tab:tnsxmatch}. 
About half of these transients are already classified and some are well-studied in the literature (e.g. SN2016bau: \citealt{Aryan2021}. SN2024ggi: \citealt{Pessi2024, JacobsonGalan2024}). For those, our HST matches may provide useful pre-outburst luminosity levels. The TNS report of AT2019ejn mentions multiple discoveries of this transient, suggesting it may be an LBV. For three other transients (AT2020aaqy, AT2021tmm and AT2021ytf), the $F814W$ magnitude is brighter than the TNS discovery magnitude, making them likely to also be variable stars. For nine other transients, the type was not straightforwardly identified in TNS and they deserve a detailed analysis of their light curves and HST progenitors. Therefore, we inspected HST colour images for every candidate, using the Hubble Legacy Archive (HLA) website\footnote{\url{https://hla.stsci.edu/hlaview.html}}. Cutouts of 10 arcsec side-to-side are displayed in Figure \ref{fig:hst_cutouts} and the position of the transient is marked by a circle. 

\begin{figure*}
    \centering
    \includegraphics[width=\textwidth]{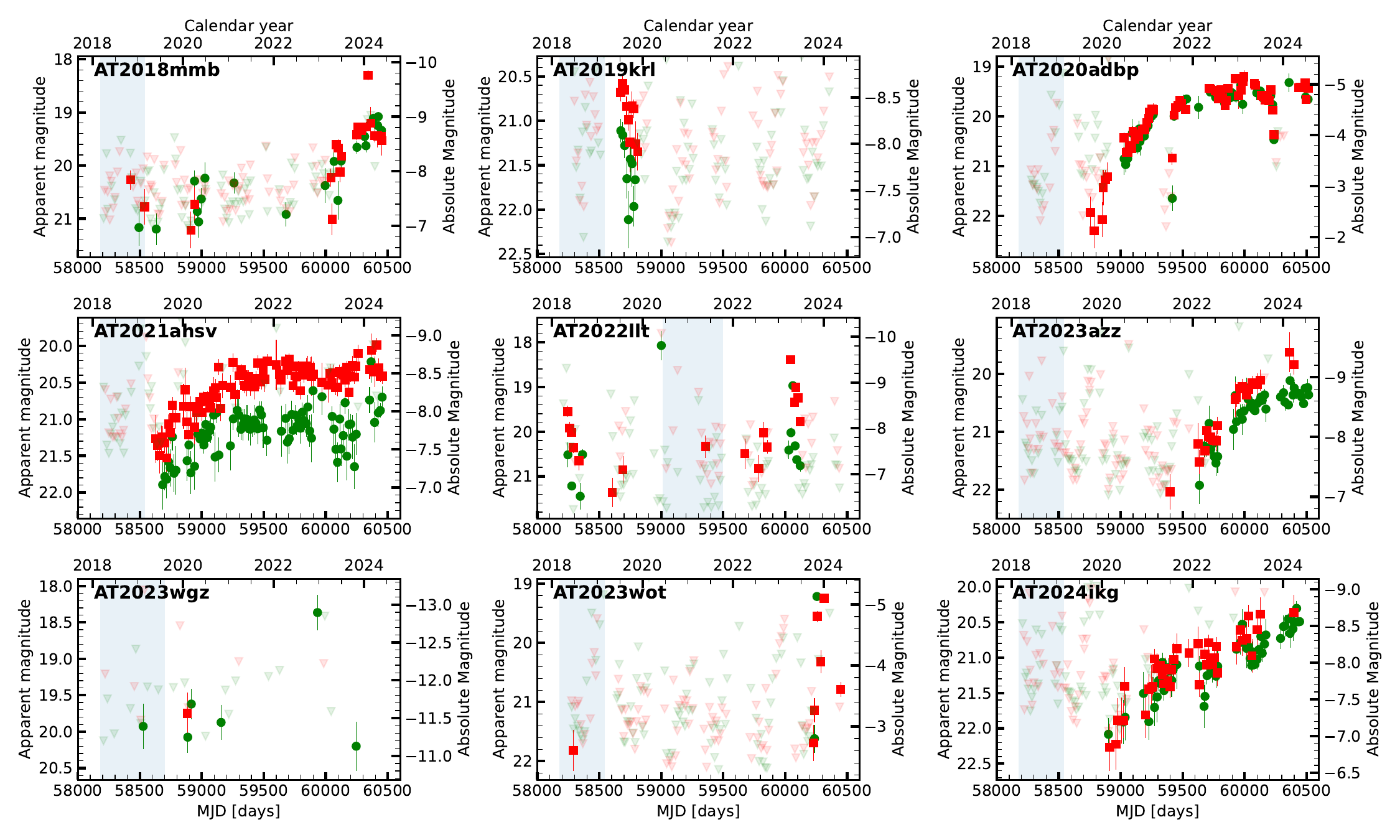}
    \caption{ZTF forced photometry light curves of the 9 candidates resulting from our TNS cross-match. The blue areas indicate the period that was used to set the baseline flux at the location of these transients. }
    \label{fig:lightcurves_candidates}
\end{figure*}

To inspect the temporal evolution of the candidates, we obtained forced photometry for the ZTF public data using the online service ZTF Forced Photometry Service (ZFPS) \citep{Masci2023arXiv}. For every candidate, we retrieved all the data starting from the beginning of the public survey in March 2018 (around MJD 58178). We generally used the difference imaging flux obtained during the first year of operations (or a period with low residual flux) as a baseline to calibrate the difference imaging magnitudes. We applied the quality cuts recommended in the documentation and we imposed \texttt{forcediffimchisq} (Reduced chi-square in PSF-fit) $<1.3$. Data points with S/N>3 were considered as detections and the upper limits are reported with a 5$\sigma$ threshold. To increase the S/N for these faint objects, we binned the fluxes using a 15-day bin size. The resulting light curves are shown in Figure \ref{fig:lightcurves_candidates}. In the following, we report the results of a detailed analysis aimed at identifying potential LRN precursors.

\textit{AT2018mmb}: this transient is located at 0.55'' from a faint ($m_{F555W}$ = 22.78, $m_{F814W}$ = 21.68) YSG in our sample, inside a stellar cluster. However, it is in a crowded region, surrounded by many sources that may cause source confusion (CI = 1.4), and a significantly brighter source ($m_{F555W}$ = 19.07, $m_{F814W}$ = 19.38) is located at just 0.37'' from the transient. The position of this bright source in the CMD is close the locus of the MS. Although this neighbour is the most likely progenitor of the transient, its recent brightening in 2024 (reaching an absolute magnitude of $-$9) makes it an interesting candidate for followup.


\textit{AT2019krl}: this transient matches a YSG candidate with a separation below 0.1'' and a stellar mass $\sim 10.3 M_\odot$. Image inspection shows that the YSG is devoid of any bright source in its immediate vicinity. AT2019krl was spectroscopically classified as SN IIn or LBV outburst by The Astronomer's Telegram, No. 12913. However, the absolute magnitude peak at $-$8.5 is several magnitudes fainter than most SN IIn. It was detected by HST at an absolute magnitude of $-$6, translating to a ZAMS stellar mass of $\sim 10 M_\odot$, less massive than most LBV. Still, their measured $H_\alpha$ line velocity ($\sim$2000 km/s) rules out a LRN-related mechanism.

\textit{AT2020adbp}: this sources matches a YSG in M31 observed by HST in January 2023, with $m_{F814W}$ = 18.7. This observation occurred after the transient report on TNS, explaining the bright HST magnitude. There is no pre-brightening image available in the HST archive, which precludes the study of the progenitor. Source confusion is unlikely for this source having CI = 1.1 in both $F814W$ and $F475W$ filters. Followup of this source is encouraged.

\textit{AT2021ahsv}: this transient matches a YSG candidate located at 0.19'' with $M_*\sim 12 M_\odot$. It is brighter than every close-by source. Standing in the middle of the Hertzsprung gap in the CMD, this candidate progenitor shows an interesting light curve with a slow brightening. Followup of this source is encouraged, as it has showed a steady brightening since 2020.



\textit{AT2022llt}: this transient matches a $> 20 M_\odot$ YSG with a separation of 0.06''. Although ZTF forced photometry does not reveal any long-term behavior, the ATLAS light curve shows a slow brightening of the source in recent years (Figure \ref{fig:prec_lc}, top-middle panel). Image inspection shows that source confusion is unlikely for this source (also supported by its CI=1.05). Followup of this source is encouraged.

\textit{AT2023azz}: this transient is located at 0.51'' from a  $\sim 9.6 M_\odot$ YSG candidate in M101, according to its TNS coordinates. The ZTF coordinates, taking advantage of a large number of epochs, are even more precise and point to a separation of 0.2''. Image inspection reveals a faint, red source close to a brighter one (at 0.75''). However, confusion may affect the photometry of this source, given the average CI of 1.4. Besides, its location on the CMD makes it also compatible with a RSG. Followup of this object is still encouraged.

\textit{AT2023wgz}: this transient has a massive ($> 20 M_\odot$) YSG counterpart at a 0.58'' separation. Its photometry may be moderately affected by source confusion, with a CI of 1.4 in both $F814W$ and $F555W$ filters. Otherwise, it is not surrounded by any source of similar or greater luminosity. Its light curve shows only a few datapoints, with no detection since 2021.

\textit{AT2023wot}: this source matches a $\sim 4 M_\odot$ YSG in M31 with a 0.58'' separation. Although it is on the faint end of our M31 selection, image inspection shows that it is not surrounded by any source of similar or greater luminosity, ruling out source confusion (CI=1.13). The Astronomer's Telegram, No. 16319 suggest it to be a nova \citep{Hornoch2023}, which is in agreement with the absolute magnitude at peak of the forced photometry light curve ($\sim -5.3$, Figure \ref{fig:lightcurves_candidates}). Such a rapid outburst also supports a nova-like behavior.

\textit{AT2024ikg}: this recent transient is found to match a bright ($m_{F555W}$ = 21.73), massive ($\sim 16 M_\odot$) YSG candidate located at 0.33'' in M101. Querying the HSCv3 detailed catalogue reveals that it was observed in Jan 2003 and Oct 2013, the only repeated filter being $F814W$ and $F435W$. The source brightened from 22.06 to 21.65 ($F814W$) and 22.62 to 21.35 (F435W) in this interval. Inspection of the image reveals a bright RSG located just 0.25'' away from the YSG, but at fainter magnitude ($m_{F555W}$ = 23.05). Besides, the CI in both $F606W$ and $F814W$ is measured between 0.95 and 1.2 in both epochs, making source confusion unlikely. Followup of this object is encouraged.

\subsection{Other precursor candidates}
\label{sec:precursors}
LRN precursors are expected to rise by only few magnitudes in several years (e.g. \citealt{Blagorodnova2020}). In this context, they may not meet the criteria to be reported to TNS (criteria may vary depending on the collaboration) until long after their detectability has begun. In order to retrieve them, we create a \textit{Lasair}\footnote{\url{https://lasair-ztf.lsst.ac.uk/}} \citep{Smith2019} watchlist to match our YSG catalogue to all ZTF transient alerts with a matching radius of 0.6'', and a \textit{Lasair} filter to keep only objects with at least 2 positive detections (\texttt{ncandgp}$\geq2$). 

To exclude variable stars, we inspected the ZTF light curve of each of the resulting 67 sources using the AlerCE broker \citep{Forster2021}, discarding all sources showing magnitudes in the past at levels brighter or similar to present levels. We used the ATLAS forced photometry tool\footnote{\url{https://fallingstar-data.com/forcedphot/queue/}} on difference images to query candidate precursors and confirm the brightening trend on longer timescales. At this stage, we obtained nine precursor candidates consistently brightening over the last few years, including four objects previously analysed in Section \ref{sec:tnsmatches} (AT2020adbp: ZTF20abhtvor, AT2021ahsv: ZTF20abbetli, AT2023azz: ZTF23aaazair, AT2024ikg: ZTF24aalfiak).

\begin{figure}
    \centering
    \includegraphics[width=8.5cm]{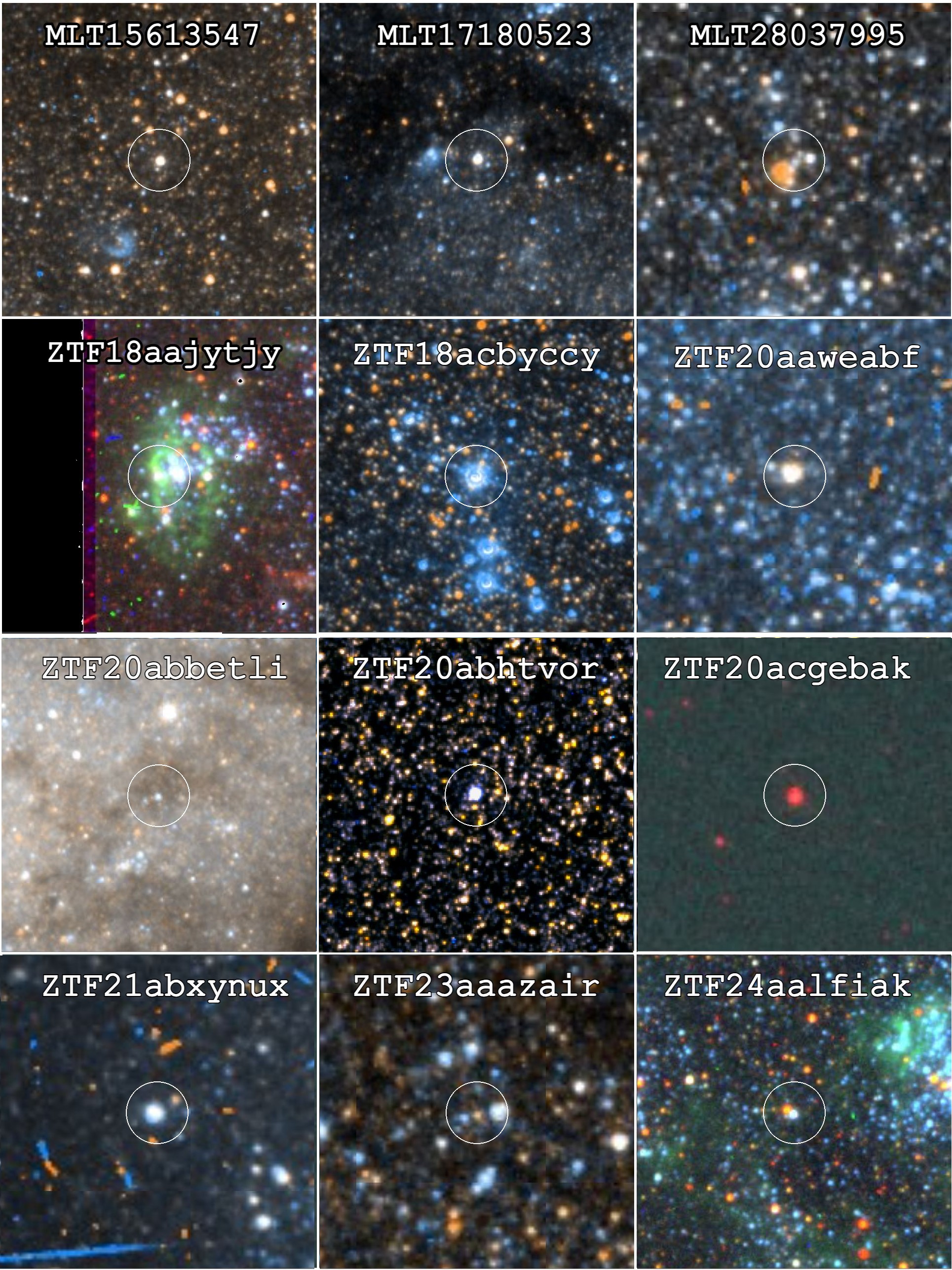}
    \caption{HST cutouts of precursor candidates. Images are 10'' side-to-side with a 1'' radius circle centered on the HST position.}
    \label{fig:prec_cutouts}
\end{figure}
\begin{figure*}
    \centering
    \includegraphics[width=\textwidth]{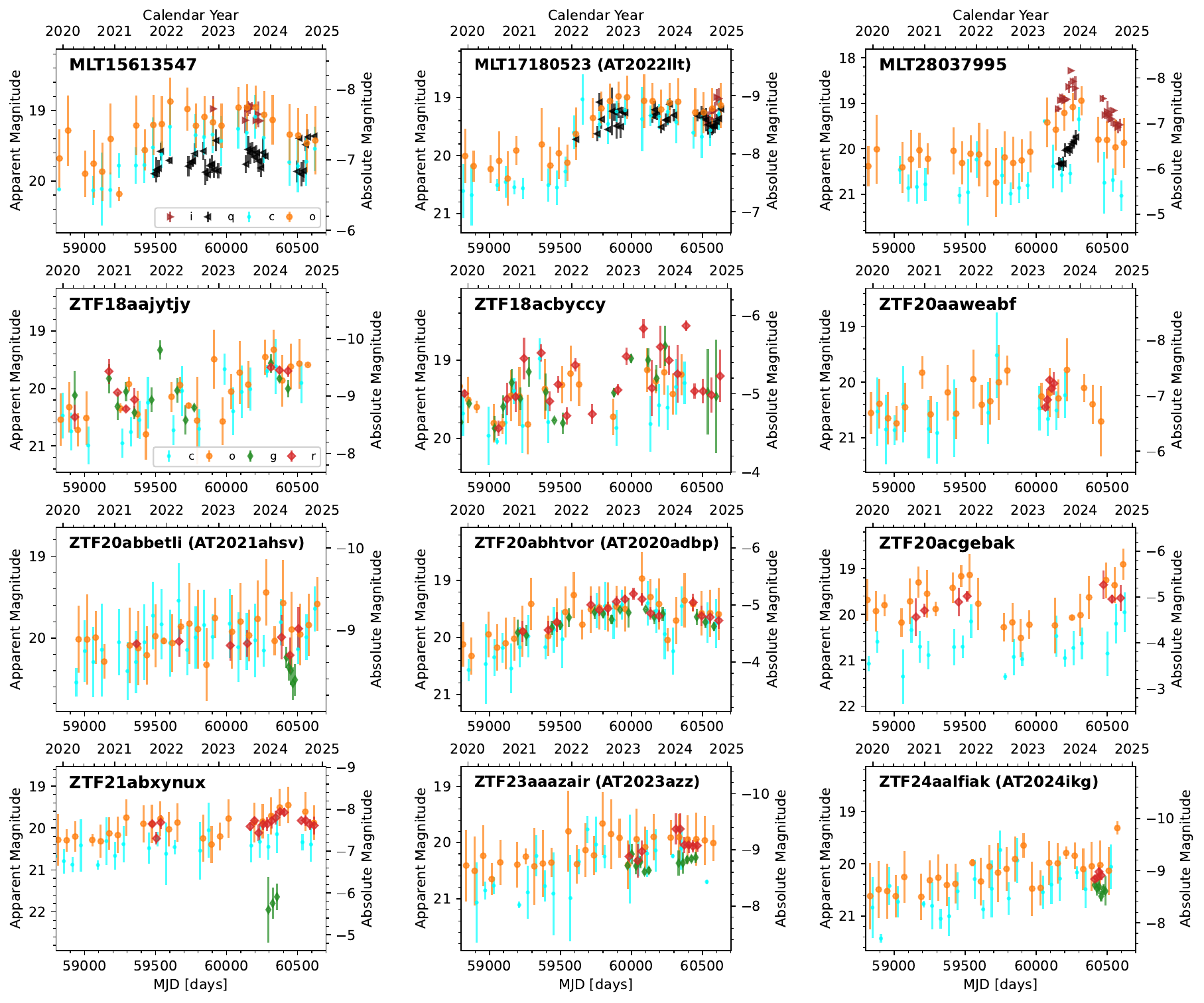}
    \caption{ATLAS forced photometry light curves of the best 12 precursor candidates in ZTF, MeerLICHT and BlackGEM. 
   ATLAS photometry is represented by orange and cyan circles ($o$ and $c$ bands), ZTF photometry by red and green diamonds ($r$ and $g$ bands) and MeerLICHT/BlackGEM photometry by black and brown triangles ($q$ and $i$ bands). ATLAS light curves were rebinned to 60-day bins, other surveys to 15-day bins.}
    \label{fig:prec_lc}
\end{figure*}

To obtain photometry for Southern candidates, we crossmatched YSG candidates to MeerLICHT and BlackGEM transients having at least 2 detections, at reasonable signal-to-noise ratio (S/N$>6.5$) and high probability to be real (rather than bogus) \texttt{class\_real}$>0.8$. The matching radius used was 1''. For BlackGEM and MeerLICHT, we obtained 10 and 33 matches, respectively. The large majority of them are variable stars with no clear brightening trend. For three of them, however, a brightening is identified and confirmed with ATLAS forced photometry: MLT28037995, in NGC 300, MLT15613547, in NGC 55, and MLT17180523 in NGC 253, actually corresponding to AT2022llt. The complete list of twelve (ZTF+MeerLICHT+BlackGEM) precursor candidates is detailed in Table \ref{tab:precursors}. Their HST cutouts are presented in Figure \ref{fig:prec_cutouts} and their multi-survey light curves are shown in Figure \ref{fig:prec_lc}. We fitted a slope to each light curve, in order to quantify the brightening trend. Slope values range between $2.1\times 10^{-4}$ and $1.2\times 10^{-3}$ mag~day$^{-1}$ , with an average error of $1.3\times 10^{-4}$ mag~day$^{-1}$. In comparison, the brightening of the precursors of M31-LRN2015 and M101 OT2015-1 were respectively of 3 mags over 2 years and 1.5 mags over 6 years \citep{Blagorodnova2017, Blagorodnova2020}, corresponding to slopes of $\sim 4\times 10^{-3}$ mag~day$^{-1}$ and $\sim 7\times 10^{-4}$ mag~day$^{-1}$, respectively. 


\begin{table}
\caption{\label{tab:precursors}Precursor candidates identified in this study.}
    \centering
    \resizebox{\columnwidth}{!}{
    \addtolength{\tabcolsep}{-0.4em}
    \begin{tabular}{cccccccc}
    \hline\hline\\ [-1.5 ex]
 Galaxy & RA & Dec & $m_{F814W}$ & Name & Sep. & Slope\\
   & (deg) & (deg) & (mag) &  & ('') & (mag/day)\\\hline\\ [-1.5 ex]
M31 & 11.04581 & 41.55485 & 17.8 & ZTF18acbyccy & 0.10 & -0.00037\\
M31 & 10.50249 & 41.14673 & 18.7 & ZTF20abhtvor & 0.01 & -0.00043\\
M33 & 23.54798 & 30.55345 & 17.7 & ZTF20acgebak & 0.03 & -0.00046\\
M101 & 210.70334 & 54.29897 & 20.8 & ZTF18aajytjy & 0.32 & -0.00037\\
M101 & 210.75242 & 54.35229 & 23.2 & ZTF23aaazair & 0.22 & -0.00036\\
M101 & 210.61563 & 54.32963 & 21.0 & ZTF24aalfiak & 0.33 & -0.00031\\
NGC 55 & 3.74938 & -39.20321 & 18.7 & MLT15613547 & 0.05 & -0.00022\\
NGC 253 & 11.97423 & -25.20828 & 19.5 & MLT17180523 & 0.28 & -0.00053\\
NGC 300 & 13.73000 & -37.60991 & 23.0 & MLT28037995 & 0.32 & -0.00083\\
NGC 1560 & 68.21523 & 71.87764 & 18.6 & ZTF21abxynux & 0.19 & -0.00054\\
NGC 6503 & 267.32443 & 70.15759 & 21.6 & ZTF20abbetli & 0.17 & -0.00037\\
UGC 9240 & 216.18222 & 44.52105 & 19.0 & ZTF20aaweabf & 0.11 & -0.00073\\

\hline
    \end{tabular}}
    \tablefoot{Precursor candidates (sorted by galaxy name) identified in ZTF, MeerLICHT, or BlackGEM. The last column gives the slope of the best-fit linear trend to the ATLAS light curve.}

\end{table}


\section{Discussion}
\label{sec:discussion}
\subsection{Comparison of our YSG candidates to spectroscopic samples}
\label{sec:spectroYSG}

The literature provides a comprehensive representation of the brightest stars in M31 and M33, as well as their nature.To assess the validity of our method and further investigate the completeness and reliability of our YSG catalogue, we compare our sample to the spectroscopic YSG and RSG samples in M31 and M33 from \cite{Drout2009}, \cite{Drout2012}, \cite{Gordon2016}, and \cite{Massey2016}. In M31, within the CMD region used to select YSGs, \cite{Drout2009} identified 120 probable YSGs and 2772 foreground dwarfs. This identification was achieved by comparing the radial velocity obtained from their spectra with the expected radial velocity at each position in M31, considering its peculiar velocity and rotation curve. Similarly, in M33, they identified 135 YSGs and 781 dwarfs \citep{Drout2012}. In a redder and less luminous region of the CMD, they identified 204 probable RSGs and 204 foreground dwarfs. The studies by \cite{Gordon2016} and \cite{Massey2016} refined these samples by identifying spectral types and members through their spectroscopic campaigns, including the extensive Local Group Galaxy Survey  (LGGS, >$1800$ spectra).

\begin{figure}
    \centering
    \includegraphics[width=8.5cm]{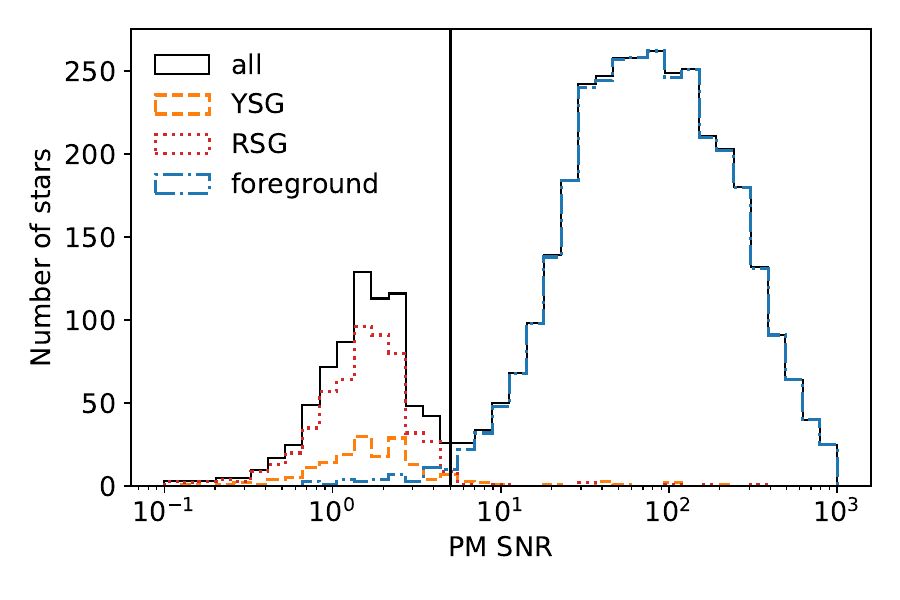}
    \includegraphics[width=8.5cm]{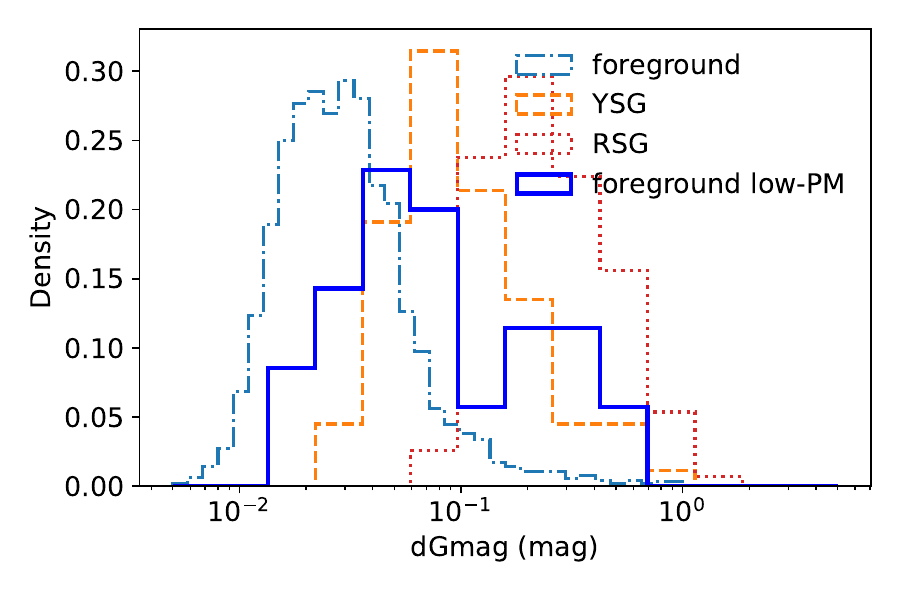}
    \caption{(Top) Distribution of the \textit{Gaia} DR3 proper motion signal-to-noise for M31 and M33 stars classified in the literature. (Bottom) Distribution of the variability indicator \texttt{dGmag} for M31 stars. The stars classified as foreground but with a low proper motion are shown in blue.}
    \label{fig:pmsnrlit}
\end{figure}

Conversely, we used \textit{Gaia} proper motions to eliminate foreground contaminants. To evaluate the effectiveness of this method, we crossmatched the samples from Drout (2009, 2012) and Gordon (2016) with \textit{Gaia} DR3. For a minority of classified dwarfs, proper motion data is unavailable, constituting an anomaly. These stars are removed from the sample. The distribution of proper motion S/N is shown in Figure \ref{fig:pmsnrlit}, categorised by the type of star they identified. The vast majority (98.5\%) of foreground sources are correctly eliminated using \textit{Gaia}'s proper motions. Additionally, 96.9\% of the eliminated sources are indeed foreground stars. \textit{Gaia} proper motions prove to be an effective method for detecting foreground stars, comparable to spectroscopy or radial velocity methods.

Comparing the samples from Drout (2009, 2012) and Gordon (2016) to our sample, we obtain 931 associations (separation less than 0.6'') for classified stars. The foreground contamination in this magnitude range is substantial, with approximately 89\% of selected stars being foreground stars. After our cleaning, 96\% of them are eliminated, and only 4\% of M31's YSGs are removed. Therefore, the completeness of our sample is preserved, leaving only 23\% residual contamination in this bright sample (typically $V<19$). At fainter magnitudes, contamination is less significant. In M33, our method performs equally good. The pre-cleaning contamination is 56\% (31 out of 56), and none of these 56 stars remain after filtering out high proper motions. 

RSGs represent another type of contaminant in our sample. This is a direct consequence of the selected regions in the CMD, which include a portion of the RGB to avoid missing YSGs affected by extinction. In literature samples, the selection region for RSGs is at a lower magnitude than for YSGs in the CMD, complicating the estimation of the fraction of RSG contaminants. At $V<18.5$, however, our selection includes 44 YSGs and 14 RSGs, resulting in a contamination rate of 24\%.

In addition to using proper motions, variability is an independent criterion that helps distinguishing a YSG from a foreground star. For example, massive YSGs have been known to vary erratically compared to low-mass dwarfs. The bottom panel of Figure \ref{fig:pmsnrlit} shows the distribution of the \textit{Gaia}-measured \texttt{dGmag}, the difference between brightest and faintest $G$-band magnitude observed during the course of the survey. Interestingly, stars classified as foreground but with a low PM exhibit variability on average 6 times greater than those with proper motion data (average \texttt{dGmag}=0.2 instead of 0.033, Figure \ref{fig:pmsnrlit}). This suggests a misclassification of these stars, which may be actual members of M31. To test this hypothesis, we model the distribution of \texttt{dGmag} as a sum of two distributions: one for high-PM sources identified as foreground, and one for low-PM sources identified as YSGs and RSGs. We can then estimate the contribution of YSGs and RSGs within low-PM foreground-classified stars. The fraction of these candidate members is 68 $\pm$ 11\%. This suggests that the residual contamination within low-PM stars with $V<18.5$ is actually $\sim$ 8\% and not 23\%. 
Furthermore, if our sample were to be used to identify progenitors of variable or transient events, the expected value of \texttt{dGmag} would be relatively large, further reducing the probability of association with a foreground source. On the other hand, although we expect most background AGN to be excluded in our sample selection (Section \ref{sec:contam}), some unresolved AGN undetected in X-rays may still produce some alerts in future deep transient surveys.

\begin{figure}
    \centering
    \includegraphics[width=8.5cm]{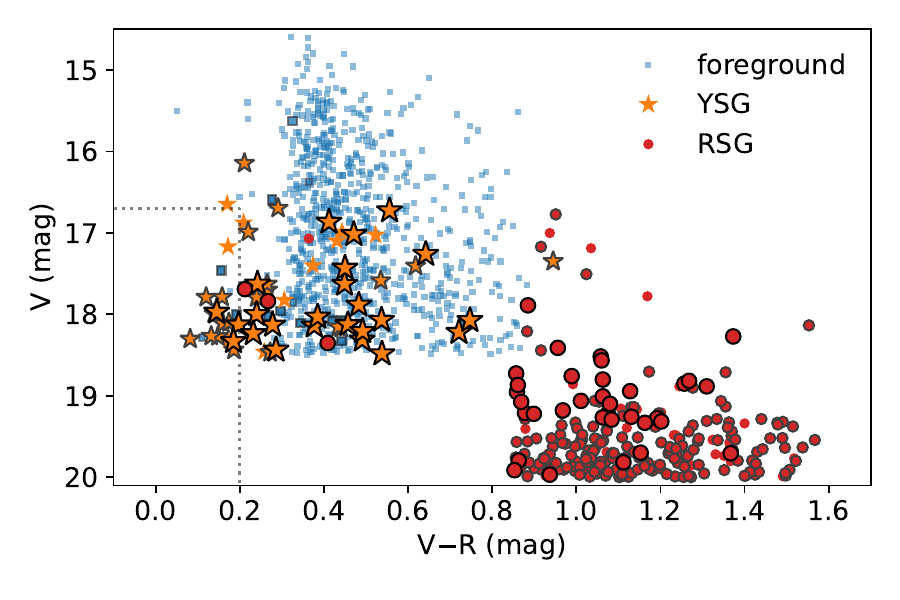}
    \caption{CMD of M31 stars classified in the literature, using the photometry of \citealt{Massey2016}. Smaller markers without contour are stars present in HST footprint but not in HST catalogues, smaller markers with grey contour are stars in HST catalogues but not in our YSG selection, and larger markers with black contour are YSG selected candidates. The dotted lines highlight a region encompassing most of HST-detected YSG that are missed in our sample.}
    \label{fig:cmds_literature}
\end{figure}

\begin{figure}
    \centering
    \includegraphics[width=8.5cm]{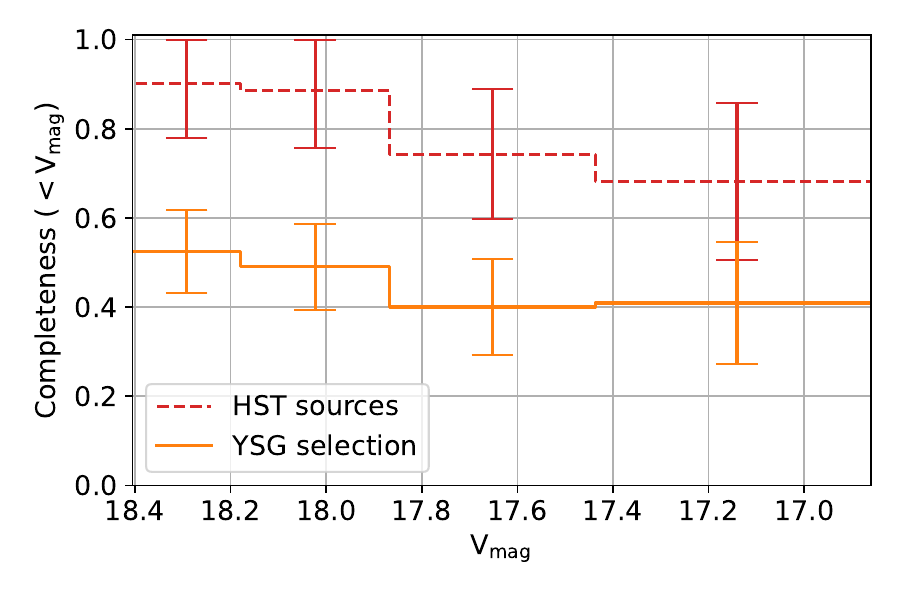}
    \caption{Fraction of M31 YSG candidates from the literature recovered as HST sources and HST-selected YSG candidates, as a function of magnitude.}
    \label{fig:completenessysg}
\end{figure}

Finally, we can quantify the completeness of our YSG catalogue in different magnitude bins. Using for reference the samples of YSG candidates in M31 from \cite{Drout2009} and \cite{Gordon2016} (see the corresponding CMD in Figure \ref{fig:cmds_literature}), and applying the same proper motion cut as done in our study, we obtain the completeness estimates shown in Figure \ref{fig:completenessysg}. While the brightest stars are not always detected in third-party HST catalogues (and surveys other than HST are well-adapted to recover those), the fraction of recovered YSGs among recovered HST sources consistently remains around 60\%. The lost fraction is primarily due to sources with high CI or located in the dotted-line rectangle shown in Figure \ref{fig:cmds_literature}. The latter occurs because our conservative threshold, designed to exclude the main-sequence and some extinguished main-sequence stars, also omits YSGs in the bluer part of the CMD.

\subsection{Impact of photometric uncertainties and source confusion}

Source confusion is commonplace in crowded fields and can significantly bias the photometry of HST sources, leading to incorrect star classifications. This happens when two stars, such as a main-sequence star and a red giant, are in close angular proximity. Their combined light can be mistaken for a single star, such as a YSG, due to the blended colour and luminosity. This effect is another source of contamination affecting our YSG catalogue.

\begin{figure}
    \centering
    \includegraphics[width=8.5cm]{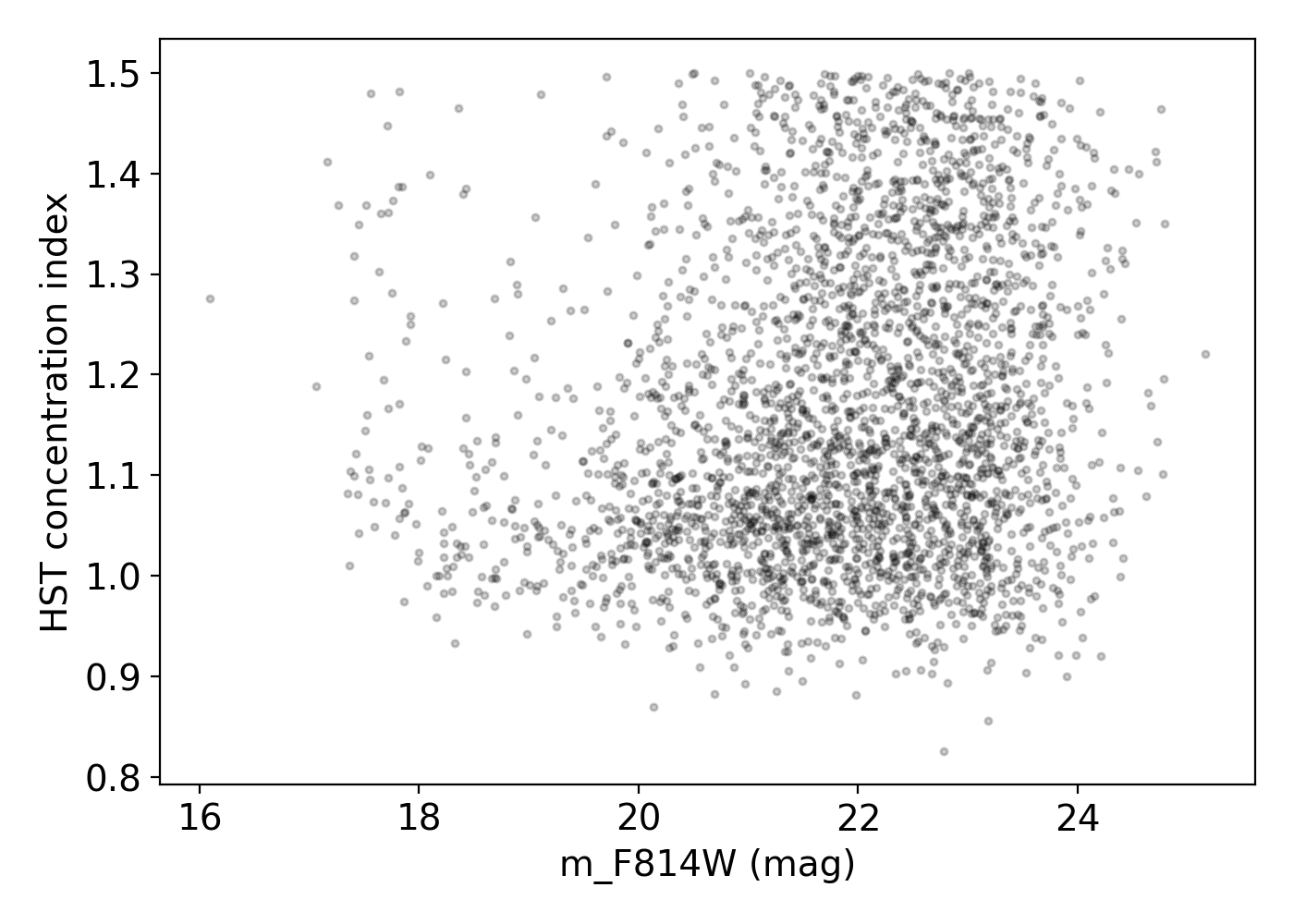}
    \caption{HST photometric concentration index of YSG candidates as a function of their $F814W$ magnitude.}
    \label{fig:cif814w}
\end{figure}

One method to quantify source confusion is by using the concentration index (CI), defined as the difference in magnitude between two apertures, typically a smaller aperture of 0.05'' and a larger aperture of 0.15'' (for ACS and WFC3/UVIS), normalised such that its distribution peaks at 1. Objects with CI values around 1.0 are likely to be stars, while those with significantly higher CI values are likely to be extended sources or the confusion of several stars. Figure \ref{fig:cif814w} presents the distribution of YSG candidates in the CI -- $m_{F814W}$ plane. At bright magnitudes, the CI is close to 1, showing that sources are point-like and source confusion is not significant. The faint sources are more numerous and thus denser on the sky, especially in distant galaxies. Therefore, they are more affected by source confusion and can have CI values close to 2. Such sources are currently excluded from our sample, which is restricted to CI values $<1.5$, thereby reducing its completeness. A future release of the catalog, incorporating DOLPHOT photometry \citep{Dolphin2016}, is anticipated to address this issue more effectively.

Similarly, photometric uncertainties can bias the selection of YSGs in HST observations. These uncertainties arise from various sources, such as photon noise, background subtraction errors, and instrumental effects. When dealing with YSG candidates, identified based on their position in the CMD, even small errors in photometry can shift stars into or out of the YSG region. For example, in the NGC 4455 galaxy, an edge-on spiral galaxy located 7.3 Mpc away, observations were conducted using the $F814W$ filter for 1030 seconds. These parameters are close to the median values of our sample of galaxies with YSG candidates. When the colour-magnitude diagram is regenerated using magnitude values and errors modelled as a normal distribution, 9\% of YSG candidates shift out of the selection region. Conversely, a similar number of previously unselected stars are shifted into the YSG selection region. Likewise, the variability of YSG can cause them to move out of the selection region; however, this variability primarily occurs over short timescales and within specific regions of the HR diagram, such as the instability strip (e.g., \citealt{Evans1993}).

In this context, our conservative selection region helps to maintain a cleaner sample, although it may exclude some legitimate YSGs. Other sources of errors, such as uncertainties in stellar models, errors in extinction, distance, or metallicity of the host, may bias the selection region to a more significant extent.

\subsection{Search for LRN Progenitors and Precursors with LSST}

In a recent paper, \cite{Strotjohann2024} explore the potential for identifying progenitor stars of core-collapse supernovae using data from ground-based wide-field surveys such as ZTF and LSST. Due to challenges like crowding and atmospheric blurring, identifying these progenitor stars in pre-explosion images is difficult. Instead, the study suggests combining numerous pre- and post-supernova images to detect the disappearance of progenitor stars. As a proof of concept, the authors implemented this approach using ZTF data. Despite analysing hundreds of images and achieving limiting magnitudes of approximately 23 in the $g$ and $r$ bands, no progenitor stars or long-lived outbursts were detected for 29 supernovae within a redshift of z $\leq$ 0.01. The sensitivity limits achieved were several magnitudes less than those in previously detected progenitors.

Conversely, the study projects that LSST, over its 10-year survey, could detect around 50 red supergiant progenitors and several yellow and blue supergiants. It estimates that progenitors of Type Ic supernovae would be detectable if they are brighter than -4.0 magnitudes in the LSST $i$ band, respectively. Given their similar spectral type (A5) compared to YSGs, we assume a similar performance for YSG detection: we therefore expect that 77\% of our YSG sample will be redetected by LSST, for sources having $M_{F814W}<-4$ (or $M_*>6M_\odot$, Figure \ref{fig:relmstarf814w}). This will provide exquisite variability constraints for sources of $r<23$, allowing us to identify new LRN precursors, given LSST repeated observations and the long timescales of these events.  Such early precursor discoveries will be essential to schedule precursor spectroscopy, allowing us to probe the line profiles, outflow structure and density (and therefore mass loss), ionization structure, and possible shocks in the system (e.g. \citealt{Pejcha2016, Molnar2017, Blagorodnova2021}).

Furthermore, LSST can expand the progenitor sample for the 1235 galaxies within 20 Mpc and at $Dec<30$º that are not covered by deep multiband HST exposures. In particular, it will cover more than 1200 massive galaxies ($M_B<-14$ to select starforming hosts), where LRNe are most likely\footnote{Every extragalactic LRN discovered in the last decade belonged to a massive starforming host, the least massive host being the Spider galaxy with $M_B=-16.7$ \citep{Pastorello2023}.}, and and whose YSGs comprise 96\% of our sample. Assuming the same distance distribution of YSG candidates as the one in this study, this represents a 130\% increase of our $M_*>6M_\odot$ sample. It is important to note that some progenitors detected only a few years before the LRN are likely already undergoing mass transfer and can be considered precursors (this is the case of e.g. AT2021blu, imaged both 15 years and 2 years prior to the transient, showing a 1-mag brightening between these two epochs \citealt{Pastorello2023}). Ongoing large-scale, deep surveys such as Euclid \citep{Euclid2011} are also of significant interest for mapping stellar populations and identifying potential transient progenitors in nearby galaxies \citep{Bonanos2024}.


The rate of LRNe in the luminosity range  $-16 \leq M_r \leq -11$ mag was recently constrained to $7.8^{+6.5}_{-3.7}\times 10^{-5}$ Mpc$^{-3}$ yr$^{-1}$ \citep{Karambelkar2023}, comparable to the core-collapse SN rate \citep{Perley2020}. However, LRNe are about three magnitudes fainter, and \cite{Karambelkar2023} suggest a luminosity function in the form ${dN}/{dL}\propto {L}^{-2.5}$. At magnitude $r<18.5$, we thus expect LSST to detect about 50 LRNe (i.e. 1.7\% of the 3300 expected bright core-collapse SNe, \citealt{Strotjohann2024}) over its 10-year survey. 

\cite{Strotjohann2024} also postulate how LSST will detect more than a thousand pre-supernova outbursts, depending on their brightness and duration. In the case of LRNe precursors, a system brightening from $-$1 to $-$4\,mag in absolute $r$-band magnitude (similar to the precursor of M31-LRN2015) would be detected in a single LSST exposure up to a distance of 5 Mpc (not taking into account the effects of the luminous background from the galaxy). This distance limit becomes 20\,Mpc when using the final survey LSST $r$-band sensitivity. Considering the host galaxy's brightness, late-time precursors or those from massive stars, with $M_r < -6$ (\citealt{Blagorodnova2020}, Figures \ref{fig:lightcurves_candidates}, \ref{fig:prec_lc}) would have a 6\% probability of detection by LSST \citep{Strotjohann2024}.  If half LRNe have such a precursor, given their volumetric rate, one can expect about 100 new precursor detections in the era of LSST. The rate, luminosity function, and timing of LRN precursors will be measurable using this large dataset. This might contribute to revealing their intrinsic mechanisms.

Overall, the cadence of LSST, its multi-band coverage, along with the depth of the survey, allows for the detection of progenitors, years-long faint precursors and variability patterns that precede LRN events. Systematic cataloging and data mining techniques will be crucial in identifying these specific observational signatures within LSST vast datasets. An approach based on the systematic use of archival data and the prediction of future variability using various light-curve analysis techniques is under study (Tranin et al., in prep.).

\section{Conclusion}

\label{sec:conclusion}

In this study, we use HST imaging of nearby galaxies to find possible LRN progenitors and precursors, making it possible to predict their outburst and rapidly identify new transients matching the position of a candidate YSG.  We retrieve the catalogues of HST sources and their photometry for 369 galaxies with distances closer than 20 Mpc, using different public databases. After building colour magnitude diagrams for each galaxy, we select the Hertzsprung gap stars using MIST stellar evolution tracks, coupled with a statistical representation of the CMD. Foreground contaminants were mostly removed using \textit{Gaia} proper motions, and the rest of foreground contaminants was quantified using the TRILEGAL simulations of the Milky Way stellar content. The previous spectroscopic identification of foreground stars within the Hertzsprung gap of M31 and M33 showed excellent agreement with our method. Additionally, we constrained the number of contaminants resulting from internal extinction to less than 20\% of the sample and quantified this for each source. The use of MIST stellar evolution tracks and a meticulous filtering process to exclude contaminants proved crucial in accurately identifying candidates.

Our study identified 146,502 yellow supergiant candidates in 353 galaxies, a significant increase over previous research. The resulting sample of candidates was cross-matched with the TNS and the ongoing surveys ZTF, BlackGEM and MeerLICHT. Candidates exhibiting outstanding variability were identified and analysed. In particular, we identified 12 precursor candidates based on their consistent brightening over the past few years. Their spectroscopic followup and identification will be the subject of an upcoming work.

The YSG catalogue resulting from this study will be released at the time of publication of this article, together with the pipeline. The Python scripts used to retrieve and analyse HST data are made publicly available at the following address: \url{https://github.com/htranin/LRNsearch}. The insights gained from this catalogue can inform future models of stellar evolution and enhance our ability to predict and study rare transient events. This work advances our capabilities to understand yellow supergiants and their role as progenitors and precursors to luminous red novae, filling a gap in previous studies. LSST will be a game changer in the quest for LRNe and their progenitors and precursors: we estimate that the 10-year survey will more than double the number of detected extragalactic YSGs within 20 Mpc, and provide excellent variability constraints for sources of magnitude $r<23$. Notably, based on the rate of previous extragalactic LRNe, we expect about 100 LRN precursors to be discovered over the course of LSST, and about 50 bright $r<18.5$ LRNe. Future research should focus on continuous monitoring of brightening YSG candidates to capture and analyse transient episodes as they occur. We emphasize the importance of closely monitoring these future transients having YSG progenitor, to ensure the identification and study of luminous red novae events and other rare transients. 

\begin{acknowledgements}
H. T. and N. B. acknowledge to be funded by the European Union (ERC, CET-3PO, 101042610). Views and opinions expressed are however those of the author(s) only and do not necessarily reflect those of the European Union or the European Research Council Executive Agency. Neither the European Union nor the granting authority can be held responsible for them. PJG is supported by NRF SARChI grant 111692. We thank Zeljko Ivezic and the anonymous referee for their valuable comments, which significantly enhanced the quality of this paper. We thank Rick White and Bernie Shiao for their assistance in retrieving the HSCv3 data. We acknowledge the extensive use of the MAST database to conduct this study.

Based on observations with the MeerLICHT telescope. MeerLICHT is built and run by a consortium consisting of Radboud University, the University of Cape Town, the South African Astronomical Observatory, the University of Oxford, the University of Manchester and the University of Amsterdam. MeerLICHT is hosted by SAAO.

Based on observations with the BlackGEM telescopes. BlackGEM is built and run by a consortium consisting of Radboud University, the Netherlands Research School for Astronomy (NOVA), and KU Leuven with additional support from Armagh Observatory and Planetarium, Durham University, Hamburg Observatory, Hebrew University, Las Cumbres Observatory, Tel Aviv University, Texas Tech University, Technical University of Denmark, University of California Davis, the University of Barcelona, the University of Manchester, University of Potsdam, the University of Valparaiso, the University of Warwick, and Weizmann Institute of science. BlackGEM is hosted and supported by ESO.
\end{acknowledgements}

\bibliographystyle{aa}
\bibliography{main}

\begin{appendix}
\begin{onecolumn}

\section{Example of CMDs}
\label{sec:appendix}
Figure \ref{fig:cmdsample} presents a representative subset of CMDs analysed in this study. The selection area of YSG candidates is shown by solid blue lines.

\label{LastPage}
\begin{figure}
    \centering
    \includegraphics[width=5.8cm]{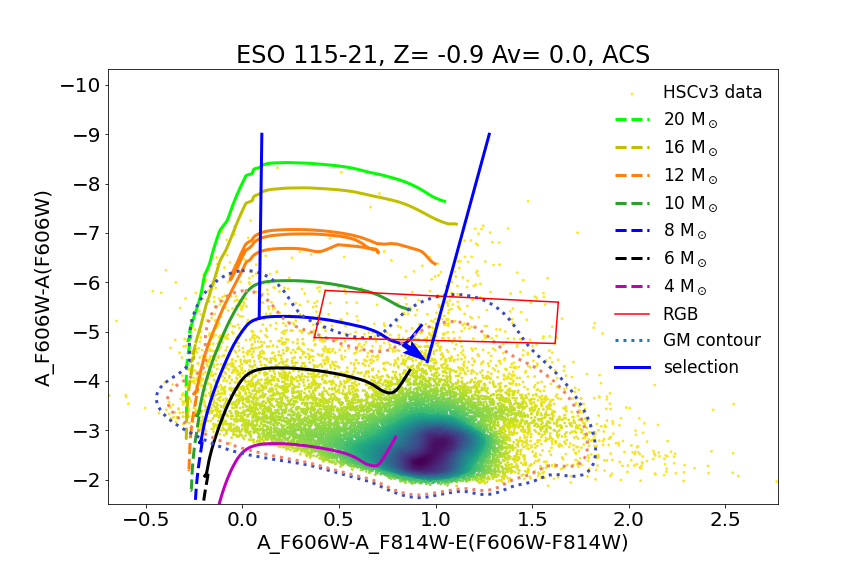}
    \includegraphics[width=5.8cm]{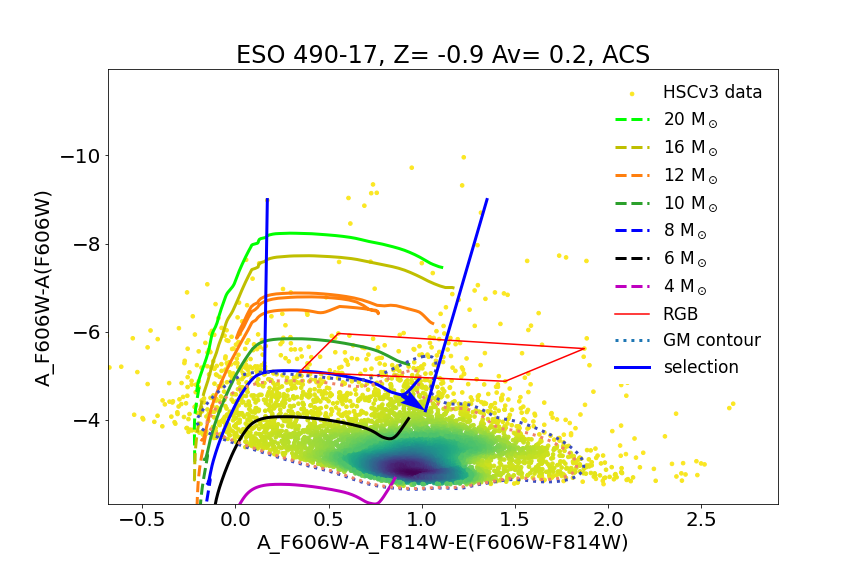}
    \includegraphics[width=5.8cm]{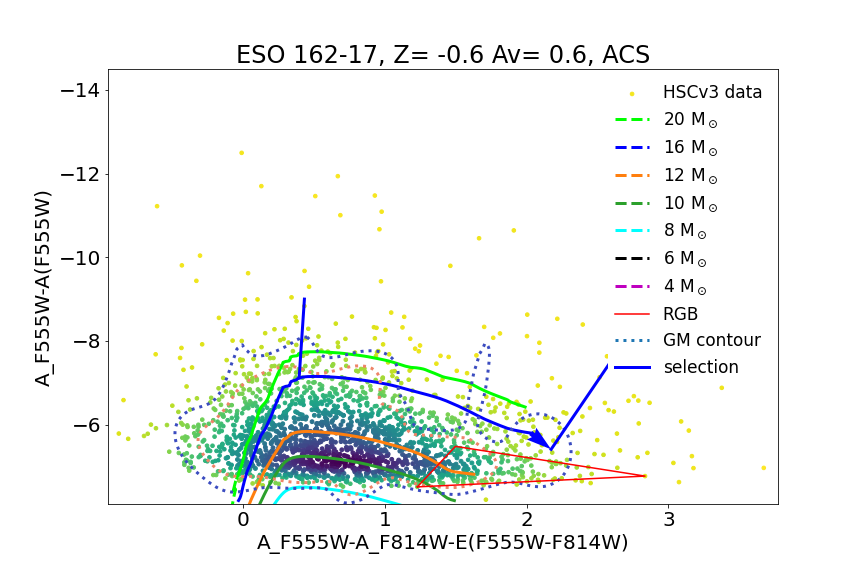}
    \includegraphics[width=5.8cm]{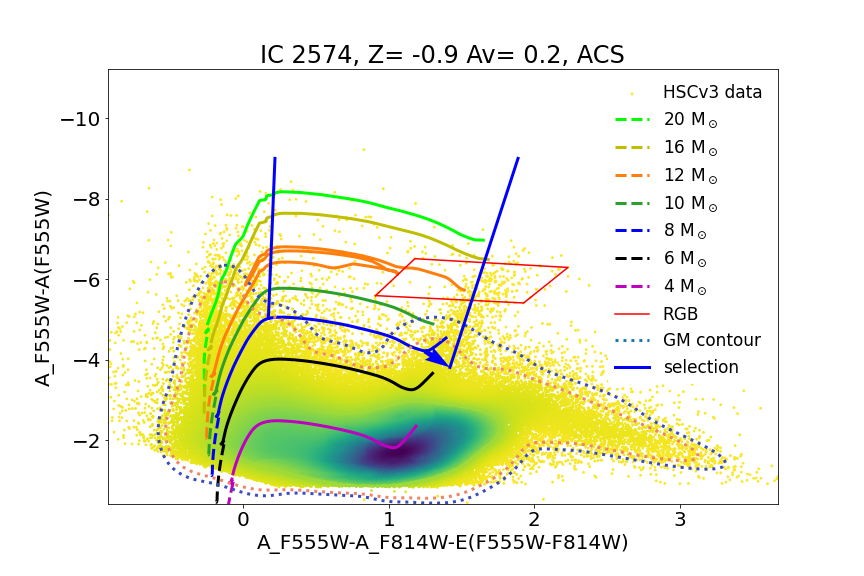}
    \includegraphics[width=5.8cm]{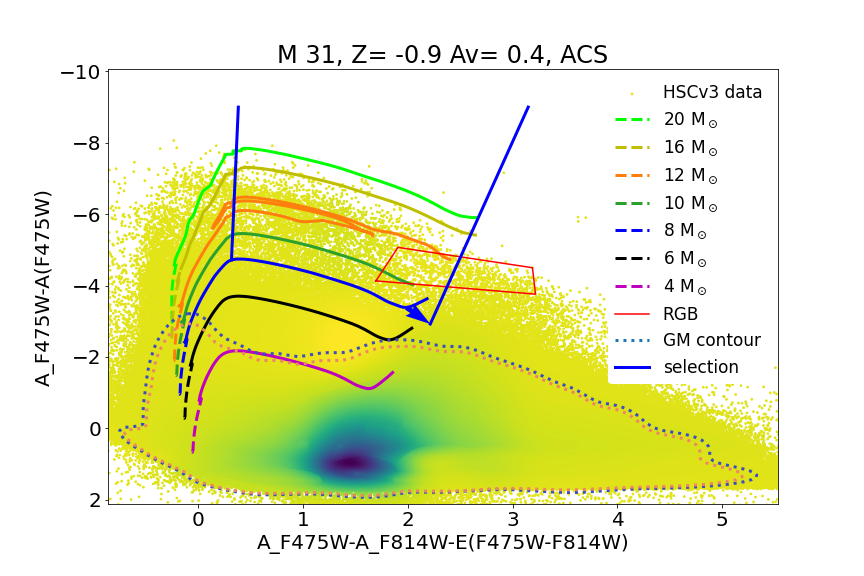}
    \includegraphics[width=5.8cm]{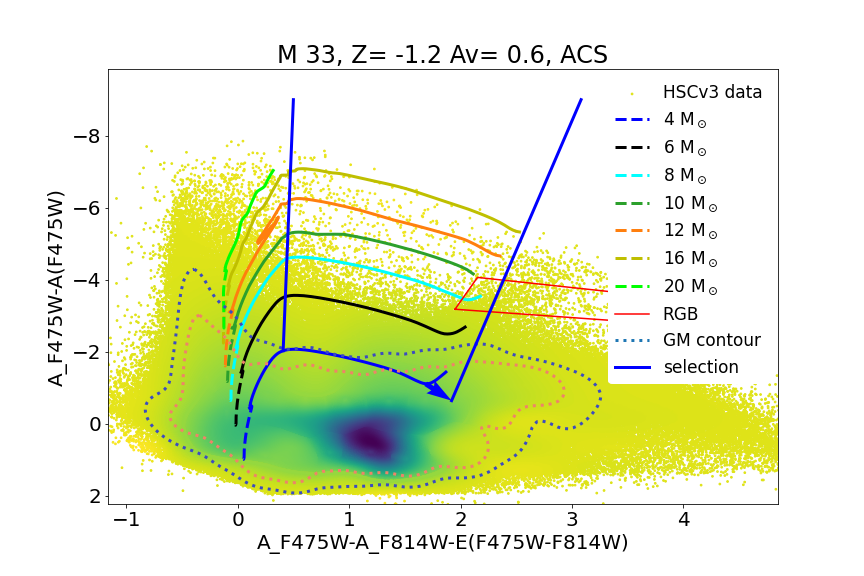}
    \includegraphics[width=5.8cm]{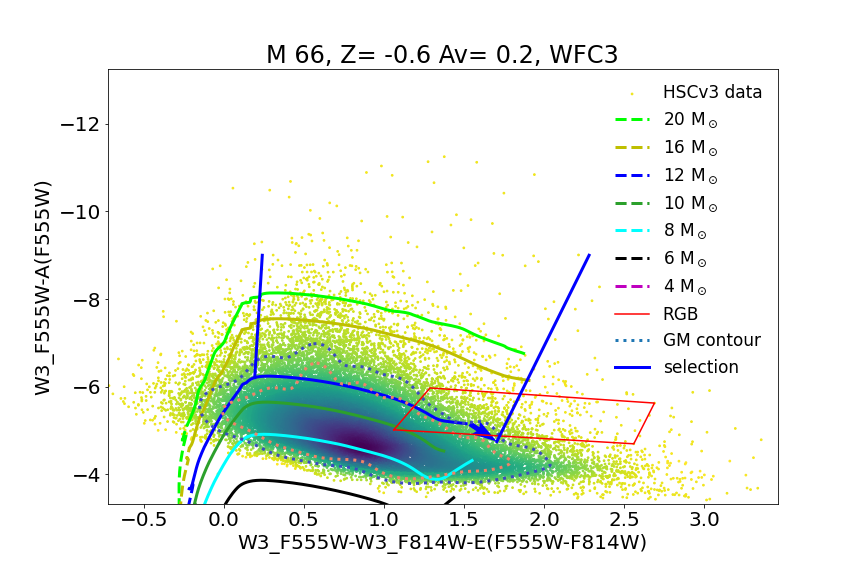}
    \includegraphics[width=5.8cm]{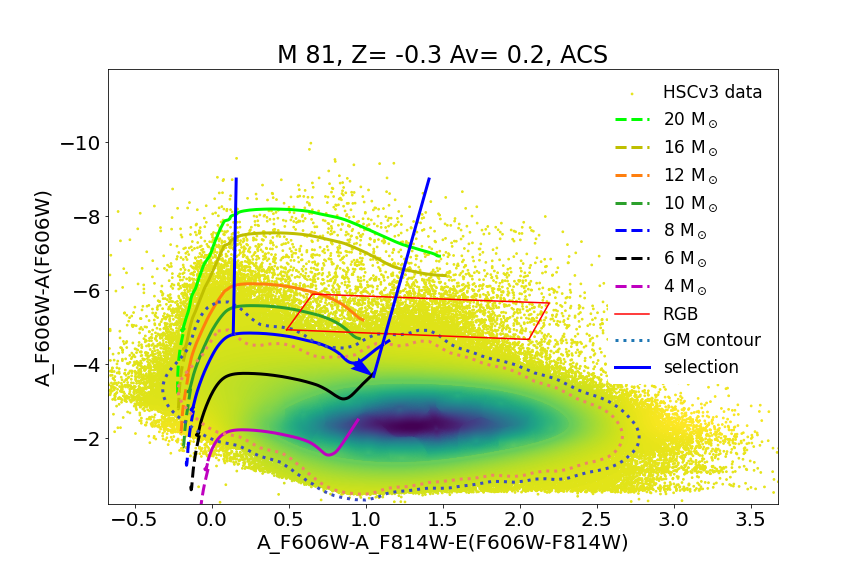}
    \includegraphics[width=5.8cm]{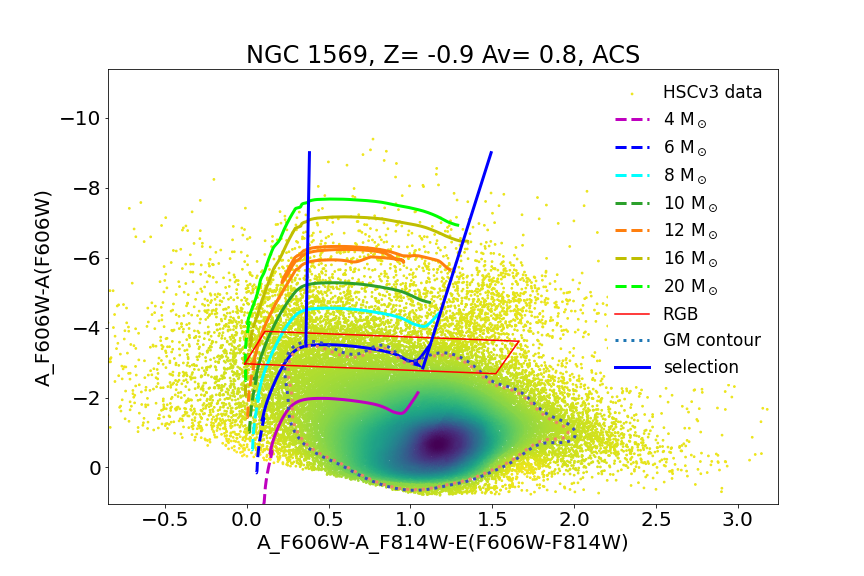}
    \includegraphics[width=5.8cm]{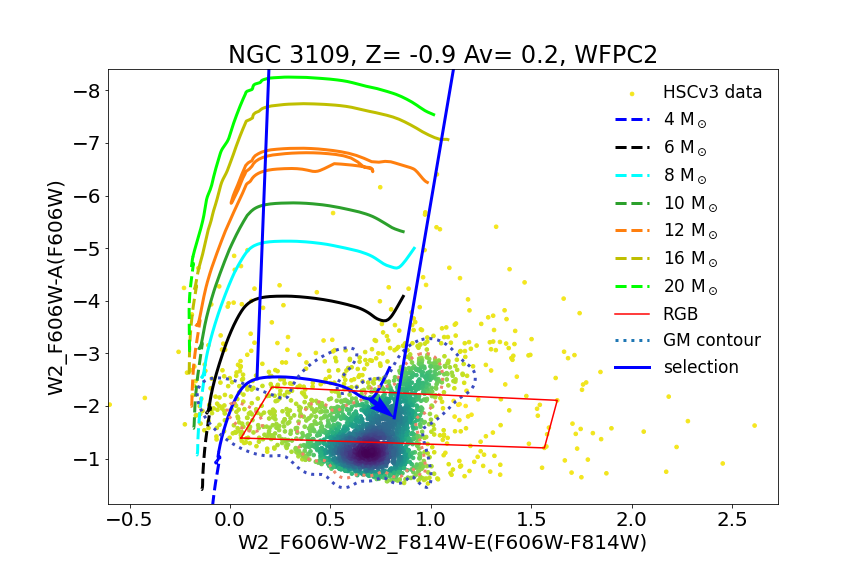}
    \includegraphics[width=5.8cm]{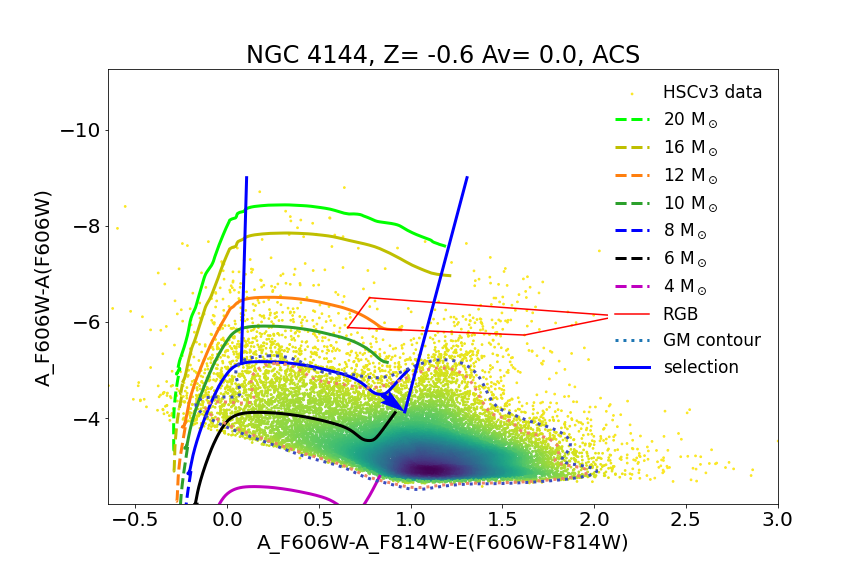}
    \includegraphics[width=5.8cm]{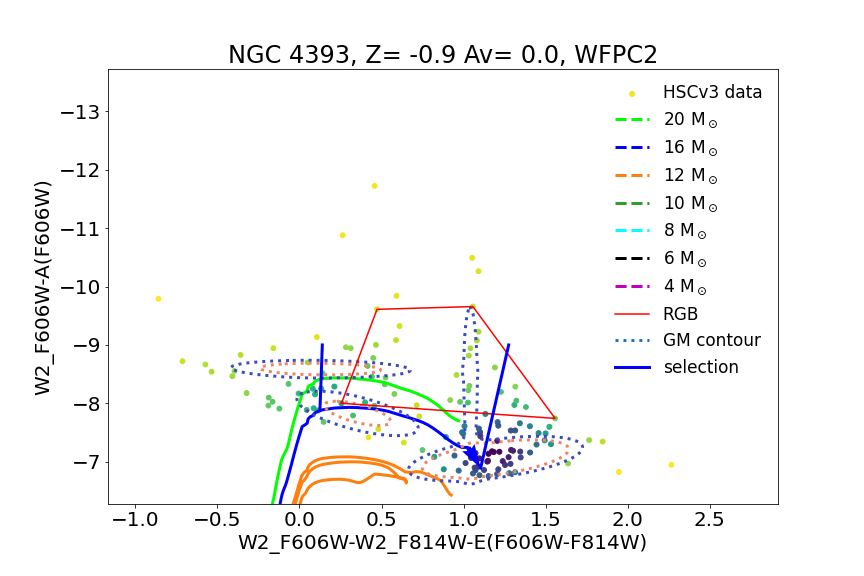}
    \includegraphics[width=5.8cm]{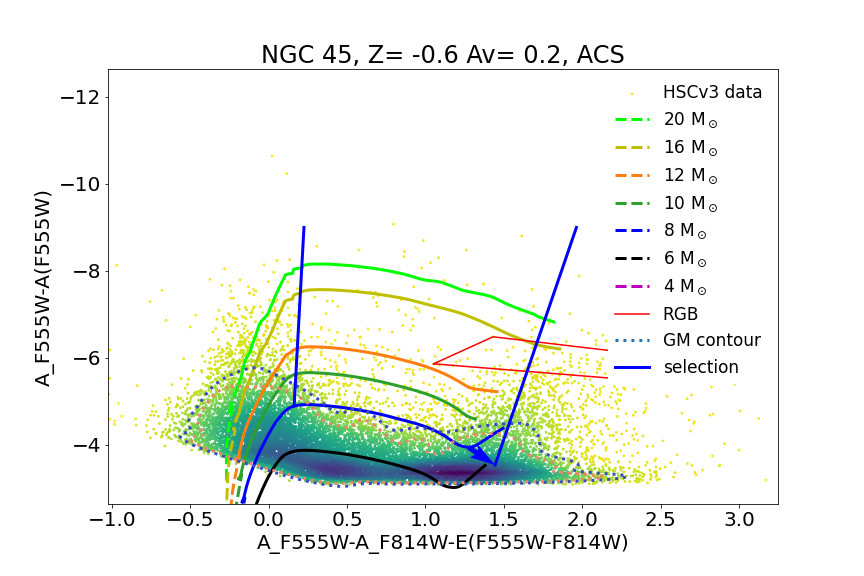}  
    \includegraphics[width=5.8cm]{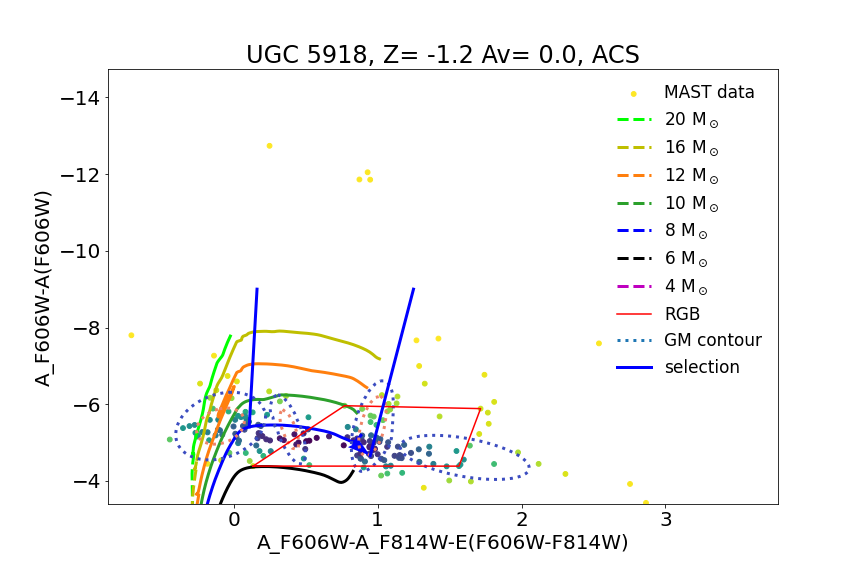}
    \includegraphics[width=5.8cm]{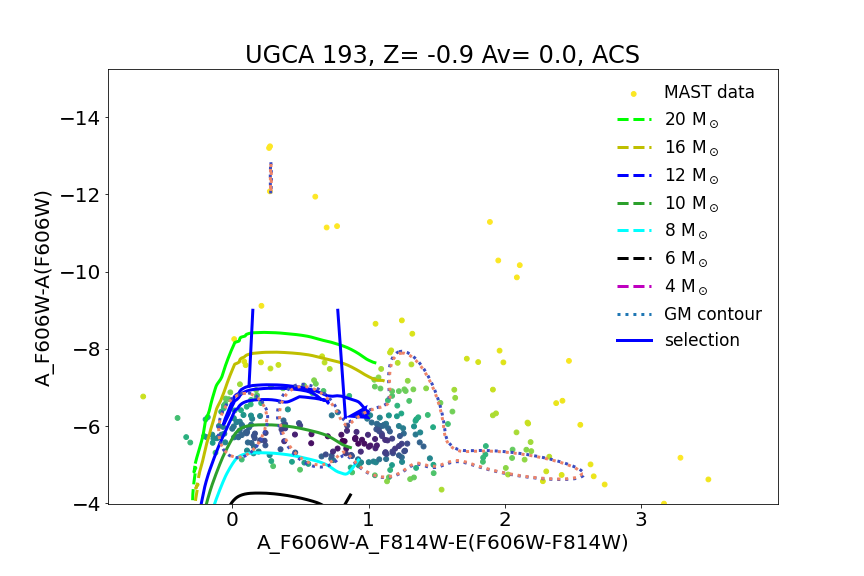}
    \caption{Representative sample of CMDs analysed in this study. The blue MIST track corresponds to the reference stellar mass at which YSG observation completeness is ensured.}
    \label{fig:cmdsample}
\end{figure}

\end{onecolumn}
\twocolumn

\end{appendix}

\end{document}